\documentclass[pra, aps, twocolumn, groupedaddress, showpacs, superscriptaddress, 10pt]{revtex4-2}
\usepackage{amssymb,amsmath,amsthm,color,times,graphicx}
\usepackage{hyperref}
\usepackage{subfigure}
\usepackage{times}
\usepackage{bbold}
\usepackage{color}

\providecommand{\openone}{\leavevmode\hbox{\small1\kern-4.3pt\normalsize1}}

\theoremstyle{plain}

\theoremstyle{definition}

\usepackage{orcidlink}
\makeatletter
\newsavebox{\@brx}
\newcommand{\llangle}[1][]{\savebox{\@brx}{\(\m@th{#1\langle}\)}%
  \mathopen{\copy\@brx\mkern2mu\kern-0.9\wd\@brx\usebox{\@brx}}}
\newcommand{\rrangle}[1][]{\savebox{\@brx}{\(\m@th{#1\rangle}\)}%
  \mathclose{\copy\@brx\mkern2mu\kern-0.9\wd\@brx\usebox{\@brx}}}
\makeatother
\begin{document}
\title{Enhancing phase sensitivity in Mach-Zehnder interferometer with various detection schemes using SU(1,1) coherent states}
\author{N.-E. Abouelkhir \orcidlink{0000-0002-6164-0525}}\affiliation{LPHE-Modeling and Simulation, Faculty of Sciences, Mohammed V University in Rabat, Rabat, Morocco.}
\author{A. Slaoui \orcidlink{0000-0002-5284-3240}}\email{Corresponding author: abdallah.slaoui@um5s.net.ma}\affiliation{LPHE-Modeling and Simulation, Faculty of Sciences, Mohammed V University in Rabat, Rabat, Morocco.}\affiliation{Centre of Physics and Mathematics, CPM, Faculty of Sciences, Mohammed V University in Rabat, Rabat, Morocco.}
\author{E. H. Saidi}\affiliation{LPHE-Modeling and Simulation, Faculty of Sciences, Mohammed V University in Rabat, Rabat, Morocco.}\affiliation{Centre of Physics and Mathematics, CPM, Faculty of Sciences, Mohammed V University in Rabat, Rabat, Morocco.}\affiliation{College of Physical and Chemical Sciences, Hassan II Academy of Sciences and Technology, Rabat, Morocco.}
\author{R. Ahl Laamara}\affiliation{LPHE-Modeling and Simulation, Faculty of Sciences, Mohammed V University in Rabat, Rabat, Morocco.}\affiliation{Centre of Physics and Mathematics, CPM, Faculty of Sciences, Mohammed V University in Rabat, Rabat, Morocco.}
\author{H. El Hadfi}\affiliation{LPHE-Modeling and Simulation, Faculty of Sciences, Mohammed V University in Rabat, Rabat, Morocco.}\affiliation{Centre of Physics and Mathematics, CPM, Faculty of Sciences, Mohammed V University in Rabat, Rabat, Morocco.}
\begin{abstract}
Improving interferometric phase sensitivity is crucial for high-precision measurements in rapidly developing quantum technologies. The Mach-Zehnder interferometer (MZI) is a versatile tool for analyzing this phenomenon. By splitting and recombining a light beam using beam splitters, MZIs allow for precise phase sensitivity analysis using tools like the quantum Cramér-Rao bound (QCRB) and the quantum Fisher information (QFI). This paper analyzes the phase sensitivity of a MZI in various scenarios using different detection schemes and input states. We compare the single- and two-parameter quantum estimation and their associated QCRB for three phase-shift situations: in both arms, only in the upper arm (asymmetric), and in both arms symmetrically. We then investigate the phase sensitivity under three detection schemes: intensity difference, single-mode intensity, and balanced homodyne. Additionally, we explore the use of Perelomov and Barut-Girardello coherent states, two types of SU(1,1) coherent states, in all scenarios. Notably, we demonstrate that under optimal conditions, all detection schemes can achieve the QCRB by utilizing SU(1,1) coherent states as input states.

\par
\vspace{0.25cm}
\end{abstract}
\date{\today}

\maketitle
\section{Introduction}
Interferometry, exploiting the interaction of superimposed waves, is essential in precision measurement, quantum metrology, and sensing applications \cite{Peters2001, Fixler2007, Abbott2016, Abbott2017}. This technique also contributes significantly to the understanding of fundamental physics concepts \cite{Ou2020}. Notably, many physical quantities, including distance, local gravity fields, and magnetic fields, are related to the phase differences of the interfering waves, highlighting the interferometer's high sensitivity to phase changes and its wide applicability in precise measurement and metrology \cite{Gerry2001, Pezze2008, Berry2009, Liu2010, Giovannetti2011, Birrittella2012, Lu2012, Gerry2012, Gagatsos2013, Slaoui2023}. Phase estimation, a cornerstone of quantum metrology, is extensively studied due to its crucial role in diverse precision applications like environmental sensing, gravitational wave detection \cite{Abbott2009, Demkowicz2013, Grote2013, Aasi2013, Oelker2014, Acernese2014, Scientific2017, Vahlbruch2018, Mehmet2018, Tse2019}, and gyroscopes \cite{Bergh1981, Li2017, Liang2017, Khial2018}. This vital role stems from the high sensitivity of optical interferometry.\par

In the conventional classical setting with standard resources, the sensitivity reaches a limit called the shot-noise limit (SNL) or the standard quantum limit (SQL), scaling as $1/\sqrt{\mathcal{N}}$, where $\mathcal{N}$ is the number of input photons \cite{Caves1981, Giovannetti2011}. To surpass this limit, Caves \cite{Caves1981} proposed the squeezed-state technique, which addresses vacuum fluctuations at the unused input port. Subsequently, various quantum resources such as entangled coherent states \cite{Xiao1987, Boto2000, Steinlechner2013}, NOON states \cite{Bollinger1996, Dowling2008}, number-squeezed states \cite{Pezze2013}, and two-mode squeezed states \cite{Anisimov2010}, have been explored to further enhance measurement precision, potentially reaching the ultimate quantum limit of $1/\mathcal{N}$, known as the Heisenberg limit \cite{Ou1997, Giovannetti2006,Ataman2020, Anisimov2010}. In quantum interferometry, the theoretical limits on phase sensitivity are derived by applying the quantum Fisher information (QFI) and its corresponding fundamental limit on precision, the quantum Cramér-Rao bound (QCRB) \cite{Braunstein1994, Demkowicz2012, Pezze2015}. Beyond their theoretical significance, these limits prove highly valuable for assessing the optimality of practical detection schemes. A novel type of interferometer, referred to as an SU(1,1) interferometer, is configured similarly to a Mach-Zehnder interferometer (MZI). Yurke et al.\cite{Yurke1986} first proposed this concept theoretically, wherein linear beam splitters (BS) are replaced by non-linear beam splitters for the coherent splitting and mixing of two input fields to achieve precise phase estimation. The term SU(1,1) is derived from the interaction type utilized in parametric processes associated with nonlinear wave mixing, distinct from the SU(2) interaction linked to linear wave mixing through a beam splitter. This nomenclature reflects the specific nature of the interaction in these two interferometer types. In this paper, we focus on the phase sensitivity of a MZI. It is recognized that most linear interferometers can be transformed into an MZI, enabling the optimization of their phase sensitivity for different input states and detection schemes. However, this transformation is not applicable to nonlinear interferometers such as the SU(1,1) setup introduced by Yurke et al.\cite{Yurke1986}, which employ non-linear beam splitters instead of linear ones.\par

A critical goal in interferometry is realizing the theoretically optimal phase sensitivity, which requires optimizing all potential estimators and detection strategies. The QFI, denoted as $\mathcal{H}$, plays a pivotal role in this context \cite{Helstrom1973, Helstrom1968,Ataman2019, Holevo1973, Wu2019, Abouelkhir2023, Ikken2023}, being directly linked to the QCRB, expressed as $\Delta\theta_{QCRB}=1/\sqrt{\mathcal{H}}$. Here, $\theta$ represents the phase shift being estimated. Therefore, finding ways to increase the QFI becomes an important issue in quantum estimation theory. The phase sensitivity ($\Delta\theta_{det}$) in any practical detection scheme always equals or exceeds the QCRB ($\Delta\theta_{QCRB}$), i.e., $\Delta\theta_{det}  \geq \Delta\theta_{QCRB}$. The phase sensitivity in a MZI is determined by the choice of input states and detection strategies. This study investigates the use of SU(1,1) coherent states (CSs) as input, aiming to improve phase estimation accuracy and achieve the fundamental sensitivity limits of quantum metrology. In previous work \cite{Abdellaoui2024}, we demonstrated that SU(2) spin-coherent states can surpass classical limits under optimized detection schemes. Building on this, the current study explores SU(1,1) coherent states to further improve phase estimation precision and approach the ultimate sensitivity bounds in quantum metrology. The SU(1,1) and SU(2) coherent states share similarities due to the close relationship between their Lie algebras. However, SU(1,1) has two relevant types: Perelomov coherent states (PCS) and Barut-Girardello coherent states (BGCS). PCS, introduced by Perelomov \cite{Perelomov1977}, are analogues of harmonic oscillator coherent states, achieved by displacing the vacuum state with a displacement operator. BGCS, introduced by Barut and Girardello \cite{Barut1971}, are defined as right eigenstates of the SU(1,1) lowering operator. This work investigates both types of SU(1,1) CS as input states, alongside three detection schemes: intensity difference, balanced homodyne detection and single-mode intensity.\par

This paper is arranged as follows: Section (\ref{Sec2}) offers a concise review of SU(1,1) CSs. Section (\ref{Sec3}) introduces conventions and a two-parameter QFI approach followed by discussion of single-parameter QFI for both asymmetric and symmetric phase shifts scenarios. In Section (\ref{Sec4}), we provide expressions for the QFIs in all three considered scenarios for input Perelomov and Barut-Girardello coherent states combined with a vacuum state. The three detection schemes are described in detail in Section (\ref{Sec5}), while Section (\ref{Sec6}) analyzes their performances with input SU(1,1) coherent states. Finally, Section (\ref{Sec7}) summarizes the work.

\section{SU(1,1) coherent states}\label{Sec2}
We begin with a brief introduction to the SU(1,1) Lie algebra. This algebra is spanned by three generators, $\hat{A}_{z}$, $\hat{A}_{+}$, and $\hat{A}_{-}$, satisfying the commutation relations
\begin{equation}
[\hat{A}_{+},\hat{A}_{-}]=-2\hat{A}_{z}, \hspace{1cm} [\hat{A}_{z},\hat{A}_\pm]=\pm \hat{A}_\pm
\end{equation}
In this work, we focus on input states in the context of optical fields, particularly the SU(1,1) CSs. To examine these states, we employ the Holstein-Primakoff realization (HPR), a potent theoretical tool in quantum optics. The SU(1,1) CSs can be characterized by a set of single-mode Bose annihilation ($\hat{b}$) and creation ($\hat{b}^{\dagger}$) operators, aligning with the HPR representation of the SU(1,1) Lie algebra. This HPR form is given by the operators
\begin{align}
&\hat{A}_{+}=\hat{b}^{\dagger}(\hat{b}^{\dagger} \hat{b}+2a)^{1/2},\notag\\&
\hat{A}_{-}=(\hat{b}^{\dagger} \hat{b}+2a)^{1/2}\hat{b},\hspace{1cm}
\hat{A}_{z}=a+\hat{b}^{\dagger}\hat{b},
\end{align}
where the operators $\hat{b}$ and $\hat{b}^{\dagger}$ satisfy the Bose algebra $[\hat{b},\hat{b}^{\dagger}]=\openone$. The Fock space in this context is constructed from the vacuum state $|a,0\rangle$ such that $\hat{b}|a,0\rangle = 0$. The Fock states $|a,g\rangle$ are then obtained by applying the creation operator $(\hat{b}^{\dagger})$ to the vacuum state ($|a,g\rangle=(\hat{b}^{\dagger})^g|a,0\rangle$). The action of the operators $\hat{A}_{z}$, $\hat{A}_{+}$, and $\hat{A}_{-}$ on the Fock space states $|a,g\rangle$ (where $g=0,1,2,...$) is given by
\begin{align}\label{action of K on Fock state} 
    \hat{A}_{+}|a,g\rangle=&\left((g+1)(2a+g)\right)^{1/2}|a,g+1\rangle,\\ \label{action of K 2} \hat{A}_{-}|a,g\rangle=&\left(g(2a+g-1)\right)^{1/2}|a,g-1\rangle,\\
 \hat{A}_{z}|a,g\rangle=&(g+a)|a,g\rangle,
\end{align} 
where $|a,0\rangle$ is the lowest normalized state. The parameter $a$, known as the Bargmann index labeling the irreducible representation,  plays a crucial role in the representation theory of the SU(1,1) Lie algebra. It determines the specific representation of the algebra and is associated with the eigenvalue determination of the SU(1,1) Casimir operator. This operator, given by 
\begin{equation}\nonumber
    \hat{C}=\hat{A}^2_z-(\hat{A}_{+} \hat{A}_{-} +\hat{A}_{-} \hat{A}_{+})/2,
\end{equation}
is a central element of the algebra, meaning it commutes with all other elements. By evaluating the eigenvalues of $\hat{C}$, we find that they have the form $a(a-1)$.\par
In the context of SU(1,1) CSs, the Bargmann index $a$ is greater than zero and takes values such as $a = 1/2, 1, 3/2, \ldots$. This index not only characterizes the representation but also influences the properties and behavior of the coherent states constructed from this algebra. Specifically, the value of $a$ affects the structure of the Fock space states $|a, g\rangle$ and the action of the SU(1,1) generators $\hat{A}_{+}$, $\hat{A}_{-}$, and $\hat{A}_{z}$ on these states. Given that the SU(1,1) group is non-compact, all its unitary representations are infinite-dimensional. This leads to the existence of different types of coherent states, such as the PCSs and the BGCSs, which are distinguished by their construction methods and their dependency on the Bargmann index. The first type consists of displacing the vacuum state by the displacement operator, which generates the PCS. The second type, the BGCS, are defined as right eigenstates of the SU(1,1) lowering operator.

\subsection{SU(1, 1) Perelomov coherent states}
Following Perelomov's work \cite{Perelomov1977}, the standard PCSs are defined as
\begin{equation}\label{1}
    |\xi,a\rangle=\hat{D}(\mu)|a,0\rangle,
\end{equation}
with $\hat{D}(\mu)$ is the displacement operator for this group, defined as
\begin{equation}
    \hat{D}(\mu)=\exp\left[\mu \hat{A}_{+} -\mu^\ast \hat{A}_{-}\right],
\end{equation}
with $\mu$ is a complex number. Using the property $\hat{A}_{+}^{\dagger}=\hat{A}_{-}$, we can show the following property of the displacement operator: $\hat{D}^+(\mu)=\hat{D}(-\mu)$. This operator $\hat{D}(\mu)$ can be rewritten as
\begin{equation}
    \hat{D}(\mu)=e^{\xi\hat{A}_{+}}e^{\eta \hat{A}_{z}}e^{-\xi^{\ast}\hat{A}_{-}},
\end{equation}
where $\xi=e^{-i\varphi}\tanh |\mu|$, $\mu=e^{-i\varphi}\vartheta/2$ and $\eta=\ln(1-|\xi|^2)$. The parameter $\vartheta$ is a hyperbolic angle with $0\le\vartheta<\infty$ and the angle $\varphi$ is azimuthal with $0\le\varphi\le 2\pi$. By using this equation, we can directly obtain the PCSs (\ref{1}) in the form of
\begin{equation} \label{PCS}
    |\xi,a\rangle=(1-|\xi|^2)^a\sum_{g=0}^{\infty}\sqrt{\frac{\Gamma(g+2a)}{g!\Gamma(2a)}}\xi^g|a,g\rangle,
\end{equation}
where $\Gamma(x)$ is the gamma function.\par

To further elucidate, the operator $\hat{D}(\mu)$ for PCSs is analogous to the usual squeezing operator in the Schwinger representation, as both are responsible for generating specific quantum states through displacement or squeezing actions. The usual squeezing operator for a single mode can be written as
\begin{equation}
S(\alpha) = \exp\left[\frac{1}{2}(\alpha b^{\dagger 2} - \alpha^{\ast} b^2)\right],
\end{equation}
where $\alpha$ is a complex parameter representing the squeezing magnitude and phase.

\subsection{Barut-Girardello Coherent States}
Here, we would like to construct the Barut-Girardello coherent state. This state is defined as the solution to the eigenvalue equation for the annihilation operator $\hat{A}_{-}$ \cite{Barut1971}:
\begin{equation}\label{kz}
    \hat{A}_{-}|\xi,a\rangle=\xi|\xi,a\rangle, \hspace{1cm} a>0,
\end{equation}
where $\xi$ is an arbitrary complex number. In addition, we can decompose the eigenstates $|\xi,a\rangle$ as a superposition of the complete orthonormal basis $\{|g,a\rangle\}$
\begin{equation}\label{decompose}
|\xi,a\rangle=\sum_{g=0}^{\infty}\langle g,a|\xi,a\rangle|g,a\rangle.
\end{equation}
Let the annihilation operator $\hat{A}_{-}$ act on Eq.(\ref{decompose}). Then, utilizing Eqs.(\ref{kz}) and (\ref{action of K 2}) along with the following orthonormality relation
\begin{equation}
    \langle g,a|g',a\rangle=\delta_{gg'}, \hspace{1cm} \sum_{g=0}^{\infty}|g,a\rangle\langle g,a|=\openone,
\end{equation}
we can obtain
\begin{equation}
     \langle g,a|\xi,a\rangle=\frac{\xi}{\sqrt{g(g+2a)}}\langle g-1,a|\xi,a\rangle.
\end{equation}
After the recurrence procedure, this equation above transforms into the following
\begin{equation}
     \langle g,a|\xi,a\rangle=\xi^g\frac{\Gamma(2a)}{g!\Gamma(g+2a)}\langle 0,a|\xi,a\rangle.
\end{equation}
When we normalize the states $|\xi,a\rangle$ to unity, we have
\begin{equation} \label{BGCS}
    |\xi,a\rangle=\sqrt{\frac{|\xi|^{2a-1}}{I_{2a-1}(2|\xi|)}}\sum_{g=0}^{\infty}\frac{\xi^g}{\sqrt{g!\Gamma(g+2a)}}|g,a\rangle,
\end{equation}
where $I_{m}$ is the modified Bessel function of order $m$, defined as
\begin{equation}
    I_{m}(x)=\sum_{n=0}^{\infty}\frac{1}{n!\Gamma(m+n+1)}\left(\frac{x}{2}\right)^{2n+m}.
\end{equation}

\section{Quantum Fisher information in Mach-Zehnder Interferometer}\label{Sec3}
\begin{widetext}

\begin{figure}[ht] 
		\includegraphics[scale=0.11]{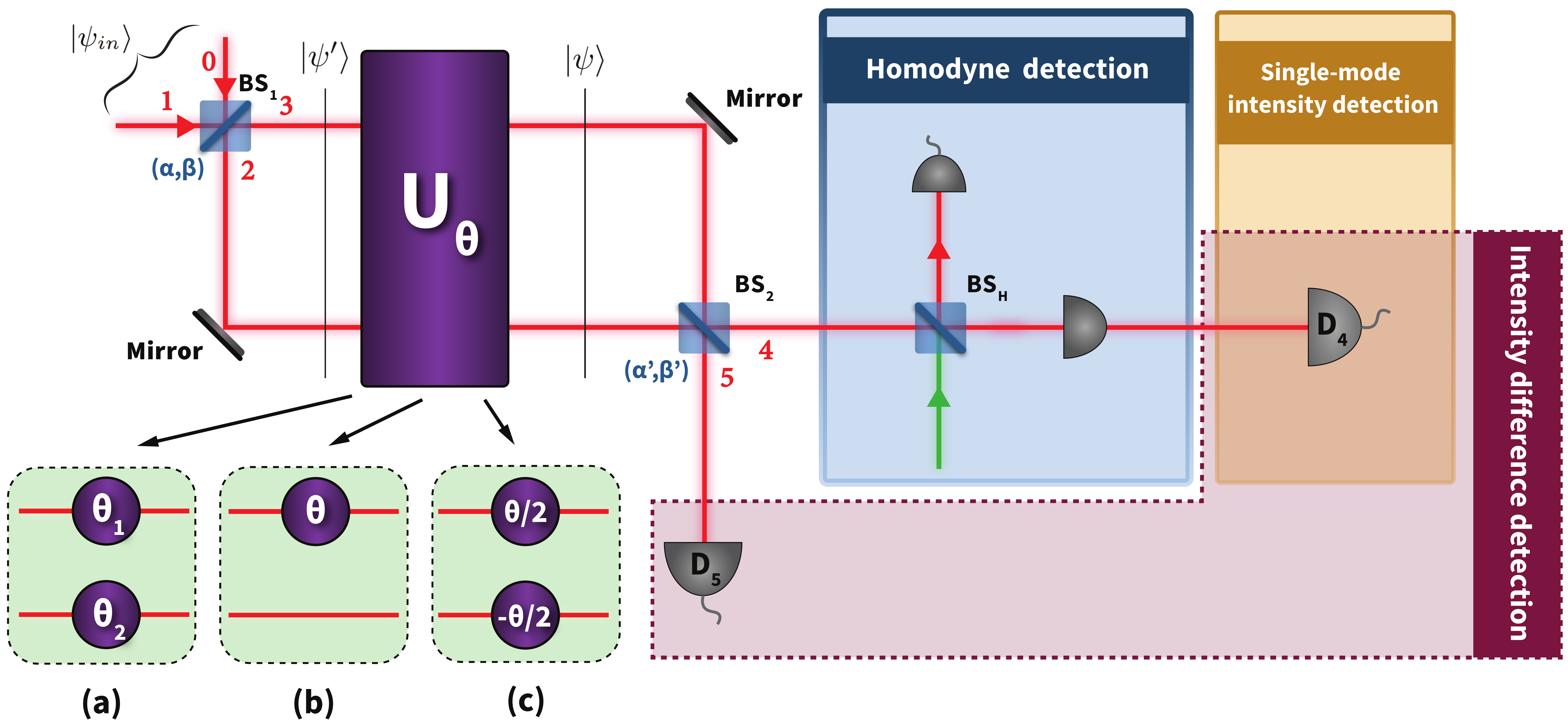} 
	\caption{Experimental setup of a Mach-Zehnder interferometer with three realistic detection schemes: intensity difference detection, single-mode intensity detection and balanced homodyne detection. In these schemes, the input state $|\psi_{in}\rangle$ undergoes a unitary transformation, resulting in the output state $|\psi_{out}\rangle$. The primary objective is to estimate the phase difference $\theta$ between the arms of the MZI using an appropriate observable.} \label{Fig1}
	\end{figure}
 \end{widetext}

We are considering the standard Mach-Zehnder (MZ) interferometric setup illustrated in Figure (\ref{Fig1}). In this setup, the two beam splitters, BS1 and BS2, have transmission (reflection) coefficients $\alpha$ ($\beta$) and $\alpha'$ ($\beta'$), respectively. Throughout this work, the input state is assumed to be pure, with no losses. In general, the precision of phase estimation in quantum interferometry is bounded by the QFI, which depends on the way the interferometer phase delay is modeled: (a) two independent phase shifts, $\theta_{1}$ ($\theta_{2}$), in the upper (lower) arm; (b)  single phase shift in the lower arm; (c) two phase shifts distributed symmetrically, $\pm\theta/2$.
 
We first consider the most general scenario, in which the upper and lower arms of an interferometer contain a phase shift, denoted by $\theta_{1}$ and $\theta_{2}$, respectively. Jarzyna and Demkowicz-Dobrzański \cite{Jarzyna2012} demonstrated effectively that relying on the single-parameter QFI often leads to overly optimistic estimates. To address this issue and avoid the problem of accounting for resources that are not actually available, it is recommended to either phase-average the input state or apply the two-parameter QFI \cite{Jarzyna2012, Lang2013,Ataman2022,Takeoka2017}.\par

Therefore, in the absence of an external phase reference, we focus on the phase shift difference $\theta_{d}=\theta_{1}-\theta_{2}$. It is more convenient to express the QFIM in the basis $\theta_{s/d}=\theta_{1}\pm\theta_{2}$. To estimate the values of $\theta_{s}$ and $\theta_{d}$, we utilize the QFIM, represented by a $2\times2$ matrix \cite{Liu2020,Bakraoui2022,Abouelkhir2023(2)}:
\begin{equation} \label{QFIM}
    \mathcal{H}=\left(\begin{array}{ll}
\mathcal{H}_{dd} & \mathcal{H}_{sd} \\
\mathcal{H}_{ds} & \mathcal{H}_{ss}
\end{array}\right),
\end{equation}
with
\begin{equation} \label{Fij}
    \mathcal{H}_{ij}=4\mathfrak{R}\{\langle\partial_{i}\psi|\partial_{j}\psi\rangle-\langle\partial_{i}\psi|\psi\rangle\langle\psi|\partial_{j}\psi\rangle\},
\end{equation}
where $\mathfrak{R}$ is the real part and the subscripts $i$ and $j$ correspond to $\theta_{s}$ and $\theta_{d}$, respectively. We consider the wavevector $|\psi\rangle$ which is expressed as
\begin{equation}
    |\psi\rangle=\exp\{-i\frac{\hat{g}_{2}-\hat{g}_{3}}{2}\theta_{d}\}\exp\{-i\frac{\hat{g}_{2}+\hat{g}_{3}}{2}\theta_{s}\}|\psi'\rangle.
\end{equation}
Here, $\hat{g}_{l}=\hat{b}^{\dagger}_{l}\hat{b}_{l}$ denotes the number operator corresponding to port $l$. We can obtain the state $|\psi'\rangle$ by applying the field operator transformations below to the input state $|\psi_{in}\rangle$;
\begin{equation} \label{transformations1}
    \hat{b}_{2}=\alpha\hat{b}_{0}+\beta\hat{b}_{1},\hspace{1cm} \hat{b}_{3}=\beta\hat{b}_{0}+\alpha\hat{b}_{1},
\end{equation}
with $|\alpha|^2+|\beta|^2=1$ and $\alpha\beta^{\ast}=-\alpha^{\ast}\beta$. This relationship implies that $\alpha^{\ast}\beta=\pm i|\alpha\beta|$. Without loss of generality, throughout this work we will adopt the convention $\alpha^{\ast}\beta=i|\alpha\beta|$. The QCRB is a lower bound on the variance of any unbiased estimator of a parameter and provides the minimum limit for estimating the parameter. In the case of multiparameter estimation, the QCRB is given as
\begin{equation}\label{QCRB}
Cov(\hat{\theta})\geq \frac{1}{N}\mathcal{H}^{-1},
\end{equation}
with $N$ repeated experiments, $\mathcal{H}$ is the QFIM given in Eq.(\ref{QFIM}) and $Cov(\hat{\theta})$ stands for the covariance matrix of the estimator, including both $\theta_{d}$ and $\theta_{s}$, whose elements are $Cov(\hat{\theta})_{ij}=E(\hat{\theta}_{i}\hat{\theta}_{j})-E(\hat{\theta}_{i})E(\hat{\theta}_{j})$ with $E$ being a mathematical expectation. Here, we set $N = 1$, and when examining phase difference sensitivity, we have
\begin{equation}
    (\Delta \theta_{d})^2\geq (\mathcal{H}^{-1})_{dd}.
\end{equation}
The expression for the first diagonal element of the inverse matrix $\mathcal{H}^{-1}$ is given by
\begin{equation}
    (\mathcal{H}^{-1})_{dd}=\frac{1}{det(\mathcal{H})}\mathcal{H}_{ss},
\end{equation}
where we can rewrite $\mathcal{H}^{(a)}$ as the reciprocal of this diagonal element:
\begin{equation} \label{F(a)}
    \mathcal{H}^{(a)}:=\frac{1}{(\mathcal{H}^{-1})_{dd}}=\mathcal{\mathcal{H}}_{dd}-\frac{(\mathcal{H}_{sd})^2}{\mathcal{H}_{ss}},
\end{equation}
thus, inequality (\ref{QCRB}) can be saturated, which implies the two-parameter QCRB is reduced to a simpler form of
\begin{equation}
\Delta \theta_{QCRB}^{(a)}= \frac{1}{\sqrt{\mathcal{H}^{(a)}}}.
\end{equation}
According to the definition of $\mathcal{H}_{ij}$ in equation (\ref{Fij}), the elements of the QFIM given in equation (\ref{QFIM}), namely $\mathcal{H}_{ss}$, $\mathcal{H}_{dd}$, and $\mathcal{H}_{sd}$, can be determined as follows
\begin{equation}
    \mathcal{H}_{ss}=\Delta^2\hat{g}_{0}+\Delta^2\hat{g}_{1},
\end{equation}
\begin{widetext}
\begin{align}
   \mathcal{H}_{dd}=&\left(2|\alpha|^2-1\right)^2\left(\Delta^2\hat{g}_{0}+\Delta^2\hat{g}_{1}\right)+8|\alpha\beta|^2\left(
\langle\hat{g}_{0}\rangle\langle\hat{g}_{1}\rangle-|\langle\hat{b}_{0}\rangle|^2|\langle\hat{b}_{1}\rangle|^2-\mathfrak{R}\left\{\langle(\hat{b}_{0}^{\dagger})^2\rangle\langle\hat{b}_{1}^2\rangle-\langle\hat{b}_{0}^{\dagger}\rangle^2\langle\hat{b}_{1}\rangle^2\right\}\right)\notag\\ &+4|\alpha\beta|^2\left(\langle\hat{g}_{0}\rangle+\langle\hat{g}_{1}\rangle\right)-8|\alpha\beta|\left(2|\alpha|^2-1\right)\left(\mathfrak{I}\left\{\left(\langle\hat{b}_{0}^{\dagger}\hat{g}_{0}\rangle-\langle\hat{b}_{0}^{\dagger}\rangle\langle\hat{g}_{0}\rangle\right)\langle\hat{b}_{1}\rangle+\langle\hat{b}_{0}\rangle\left(\langle\hat{b}_{1}^{\dagger}\hat{g}_{1}\rangle-\langle\hat{g}_{1}\rangle\langle\hat{b}_{1}^{\dagger}\rangle\right)\right\}\right), 
\end{align}
\begin{equation}
   \mathcal{H}_{sd}=\mathcal{H}_{ds}=\left(2|\alpha|^2-1\right)\left(\Delta^2\hat{g}_{0}-\Delta^2\hat{g}_{1}\right)+4|\alpha\beta|\mathfrak{I}\left\{\langle\hat{b}_{0}\rangle\langle\hat{b}^{\dagger}_{1}\rangle+\left(\langle\hat{g}_{0}\hat{b}_{0}\rangle-\langle\hat{g}_{0}\rangle\langle\hat{b}_{0}\rangle\right)\langle\hat{b}^{\dagger}_{1}\rangle+\langle\hat{b}_{0}\rangle\left(\langle\hat{b}_{1}^{\dagger}\hat{g}_{1}\rangle-\langle\hat{b}^{\dagger}_{1}\rangle\langle\hat{g}_{1}\rangle\right)\right\},
\end{equation} 
\end{widetext}
where $\Delta^{2}\hat{g}$ represents the variance of the number operator $\hat{g}$, defined as $\Delta^{2}\hat{g}=\langle\hat{g}^{2}\rangle-\langle\hat{g}\rangle^{2}$.\par

In the case of a phase shift in one arm, assuming it occurs in output $3$ of BS1, and using the notations from Fig.(\ref{Fig1}), we have the state transformed as $|\psi\rangle=e^{-i\theta\hat{g}_{3}}|\psi'\rangle$. Using definition (\ref{Fij}), the single-parameter QFI, denoted as $\mathcal{H}^{(b)}$, is given by
\begin{equation}
   \mathcal{H}^{(b)}= 4\Delta^2\hat{g}_{3},
\end{equation}
and the QCRB, which provides the optimal phase estimation, is given by
\begin{equation}
\Delta \theta_{QCRB}^{(b)}= \frac{1}{\sqrt{4\Delta^2\hat{g}_{3}}}.
\end{equation}
By applying the field operator transformations (\ref{transformations1}), we obtain the following expression
{\small
\begin{align} \label{F(b)}
\mathcal{H}^{(b)}=&4|\beta|^4\Delta^2\hat{g}_{0}+4|\alpha|^4\Delta^2\hat{g}_{1}\\
&+4|\alpha\beta|^4\left(\langle\hat{g}_{0}\rangle+\langle\hat{g}_{1}\rangle+2\left(\langle\hat{n}_{0}\rangle\langle\hat{n}_{1}\rangle-|\langle\hat{n}_{0}\rangle|^2|\langle\hat{n}_{1}\rangle|^2\right)\right)\notag\\
&-8|\alpha\beta|^2\mathfrak{R}\left\{\langle\hat{b}^2_{0}\rangle\langle(\hat{b}^{\dagger}_{1})^2\rangle-\langle\hat{b}_{0}\rangle\langle\hat{b}^{\dagger}_{1}\rangle\right\}-8|\alpha\beta|\mathfrak{I}\left\{\langle\hat{b}_{0}\rangle\langle\hat{b}^{\dagger}_{1}\rangle\right\}\notag\\ \nonumber
&-16|\alpha\beta||\beta|^2\mathfrak{I}\left\{\left(\langle\hat{g}_{0}\hat{b}_{0}\rangle-\langle\hat{g}_{0}\rangle\langle\hat{b}_{0}\rangle\right)\langle\hat{b}^{\dagger}_{1}\rangle\right\}\\ 
&-16|\alpha\beta||\alpha|^2\mathfrak{I}\left\{\langle\hat{b}_{0}\rangle\left(\langle\hat{b}^{\dagger}_{0}\hat{g}_{1}\rangle-\langle\hat{g}_{1}\rangle\langle\hat{b}^{\dagger}_{1}\rangle\right)\right\}. \nonumber
\end{align}}
Comparing equation (\ref{F(b)}) with the elements of the QFIM derived from the first estimation scenario, we can see that the single-parameter QFI, $\mathcal{H}^{(b)}$, can be expressed in terms of the coefficients of the QFIM as follows
\begin{equation}
\mathcal{H}^{(b)}=\mathcal{H}_{dd}+\mathcal{H}_{ss}-2\mathcal{H}_{sd}.
\end{equation}
If $\mathcal{H}_{sd} = \mathcal{H}_{ss}$, then the above equation and the two-parameter QFI (\ref{F(a)}) are equal, i.e., $\mathcal{H}^{(a)}=\mathcal{H}^{(b)}$. In this case, we can prove that
\begin{equation}
    \mathcal{H}^{(b)} \ge\mathcal{H}^{(a)}.
\end{equation}
In the last case, which is essentially a single-parameter estimation problem similar to the second scenario, it is modeled by the unitary operation as $U(\theta) = e^{i\frac{\theta}{2}(\hat{g}_{2}-\hat{g}_{3})}$. Consequently, the QFI is given by
\begin{equation}
   \mathcal{H}^{(c)}= \Delta^2\hat{g}_{2}+\Delta^2\hat{g}_{3}.
\end{equation}
Similarly in equation (\ref{F(b)}), we obtain
{\small
\begin{align} \label{F(c)}
&\mathcal{H}^{(c)}=\left(|\alpha|^4+|\beta|^4\right)\left(\Delta^2\hat{g}_{0}+\Delta^2\hat{g}_{1}\right)\\\nonumber
&+2|\alpha\beta|^2\left(\langle\hat{g}_{0}\rangle+\langle\hat{g}_{1}\rangle+2\left(\langle\hat{g}_{0}\rangle\langle\hat{g}_{1}\rangle-|\langle\hat{b}_{0}\rangle|^2|\langle\hat{b}_{1}\rangle|^2\right)\right)\\\nonumber
&-2|\alpha\beta|^2\left(\langle\hat{b}^{2}_{0}\rangle\langle(\hat{b}^{\dagger}_{1})^{2}\rangle+\langle(\hat{b}^{\dagger}_{0})^{2}\rangle\langle\hat{b}^{2}_{1}\rangle-\langle\hat{b}_{0}\rangle^2\langle\hat{b}^{\dagger}_{1}\rangle^2-\langle\hat{b}^{\dagger}_{0}\rangle^2\langle\hat{b}_{1}\rangle^2\right)\\\nonumber
&+2\alpha^{\ast}\beta\left(2|\alpha|^2-1\right)\left(\langle\hat{b}^{\dagger}_{0}\hat{g}_{0}\rangle-\langle\hat{b}^{\dagger}_{0}\rangle\langle\hat{g}_{0}\rangle\right)\langle\hat{b}_{1}\rangle\\\nonumber
&-2\alpha^{\ast}\beta\left(2|\alpha|^2-1\right)\left(\langle\hat{g}_{0}\hat{b}_{0}\rangle-\langle\hat{g}_{0}\rangle\langle\hat{b}_{0}\rangle\right)\langle\hat{b}^{\dagger}_{1}\rangle\\\nonumber
&+2\alpha^{\ast}\beta\left(2|\alpha|^2-1\right)\langle\hat{b}_{0}\rangle\left(\langle\hat{b}^{\dagger}_{1}\hat{g}_{1}\rangle-\langle\hat{b}^{\dagger}_{1}\rangle\langle\hat{g}_{1}\rangle\right)\\\nonumber
&-2\alpha^{\ast}\beta\left(2|\alpha|^2-1\right)\langle\hat{b}^{\dagger}_{0}\rangle\left(\langle\hat{g}_{1}\hat{b}_{1}\rangle-\langle\hat{g}_{1}\rangle\langle\hat{b}_{1}\rangle\right),
\end{align}
}
and this implies that the corresponding QCRB becomes
\begin{equation}
\Delta \theta_{QCRB}^{(c)}= \frac{1}{\sqrt{\mathcal{H}^{(c)}}}.
\end{equation}

\section{Phase estimation with a vacuum state in the first input and SU(1,1) coherent states in the second input}\label{Sec4}
In this section, our focus lies on input states characterized by SU(1,1) coherent states combined with a vacuum state. We have identified two principal types of SU(1,1) CSs: The first type, termed as the PCSs (see Eq.(\ref{PCS})), and the second type, the BGCSs (see Eq.(\ref{BGCS})). The input state is represented as follows
\begin{equation}\label{input state}
|\psi_{in}\rangle=|\xi_{i},a\rangle_{1}\otimes|0\rangle_{0},
\end{equation}
where the subscript $i=P$ or $B$ corresponds to PCS (Eq.\ref{PCS}) and BGCS (Eq.\ref{BGCS}), respectively. We also denote the QFI as $\mathcal{H}_{i}$. Using the results of the three QFIs reported in the previous section, i.e., $\mathcal{H}^{(a)}$, $\mathcal{H}^{(b)}$, and $\mathcal{H}^{(c)}$, given by equations (\ref{F(a)}), (\ref{F(b)}), and (\ref{F(c)}), it is easy to verify that the QFIs in our input states are given by
\begin{align}\label{H(a)}
\mathcal{H}^{(a)}_{i}=&4|\alpha\beta|^{2}\langle\hat{g}_{1}\rangle_{i},\\ \label{H(b)}
\mathcal{H}^{(b)}_{i}=&4|\alpha|^{4}\Delta^{2}\hat{g}_{1}+4|\alpha\beta|^{2}\langle\hat{g}_{1}\rangle_{i},\\\label{H(c)}
\mathcal{H}^{(c)}_{i}=&(|\alpha|^{4}+|\beta|^{4})\Delta^{2}\hat{g}_{1}+2|\alpha\beta|^{2}\langle\hat{g}_{1}\rangle_{i}.
\end{align}
Interferometric phase sensitivity depends on the phase difference between the two arms of an interferometer. Typically, the best sensitivity is achieved when the interferometer is balanced,  as demonstrated in \cite{Lang2013, Lang2014}. However, there are exceptions to this; for example, the works in \cite{Takeoka2017, Preda2019} shows that certain unbalanced configurations can still achieve high sensitivity. When deviating from the balanced case to the extremes where $|\alpha|$ approaches 1 or 0, it is generally assumed that interferometric phase sensitivity is lost, unless an external phase reference is used.\par
In fact, we can determine the optimal transmission coefficient ($\alpha_{\text{opt}}$) for the first beam splitter by maximizing the QFI in each scenario. For the case of a phase shift in one arm, given $\langle \hat{b}_0 \rangle = 0$ and $\langle \hat{b}_0^2 \rangle = 0$ (when port 0 is in the vacuum state), we obtain
\begin{equation}
    \mathcal{H}^{(b)} = 4\langle \hat{g}_1 \rangle |\alpha|^2 - 4(\langle \hat{g}_1 \rangle - \Delta^2 \hat{g}_1)|\alpha|^{4}.
    \end{equation}
Assuming $\alpha$ is real, we conclude that when $\Delta^2 \hat{g}_1 \ge \frac{\langle \hat{g}_1 \rangle}{2}$, the optimal transmission coefficient is $\alpha_{\text{opt}}^{(b)} = 1$, leading to the maximum QFI;
\begin{equation}
   \mathcal{H}^{(b)} = 4\Delta^2 \hat{g}_1.
\end{equation}
However, if $\Delta^2 \hat{g}_1 < \frac{\langle \hat{g}_1 \rangle}{2}$, the optimal transmission coefficient is given by
\begin{equation}
    \alpha_{\text{opt}}^{(b)} = \sqrt{\frac{\langle \hat{g}_1 \rangle}{2(\langle \hat{g}_1 \rangle - \Delta^2 \hat{g}_1)}}.
\end{equation}
In this case, the maximum QFI becomes
\begin{equation}
    \mathcal{H}_{\text{max}}^{(b)} = \frac{\langle \hat{g}_1 \rangle^2}{\langle \hat{g}_1 \rangle - \Delta^2 \hat{g}_1}.
\end{equation}
We can establish that $\mathcal{H}^{(b)}\geq \mathcal{H}^{(a)}$, indicating an advantage for quantum metrology when using an external phase reference. A crucial question arises: under what conditions do input states render the external phase reference superfluous, regardless of the transmission coefficient $\alpha$? We explore this using the asymmetric single-parameter QFI. The condition where the external phase reference adds no metrological benefit is given by $\mathcal{H}^{(b)}=\mathcal{H}^{(a)}$, and this must be true for any $\alpha$.\par
Using the equations (\ref{H(a)}), (\ref{H(b)}), and (\ref{H(c)}) derived above , we can calculate the QFI for each type of SU(1,1) coherent state.  We denote the QFI for the first type PCSs as $\mathcal{H}_{P}$ and the QFI for the second type BGCSs as $\mathcal{H}_{B}$. For PCSs, the analytical expressions for the QFIs simplify to
{\small
\begin{align}
\mathcal{H}^{(a)}_{P}=&4a|\alpha\beta|^{2}\left(\cosh(v)-1\right),\\
\mathcal{H}^{(b)}_{P}=&4a|\alpha|^{2}\left(\frac{1}{2}|\alpha|^2\sinh^2(v)+|\beta|^2\left(\cosh(v)-1\right)\right),\\
\mathcal{H}^{(c)}_{P}=&\frac{a}{2}\left(|\alpha|^4+|\beta|^4\right)\sinh^2(v)+2a|\alpha\beta|^2\left(\cosh(v)-1\right).
\end{align}
}
Therefore, the corresponding QCRBs are
{\small
\begin{align}
\Delta\theta^{(a)}_{QCRB,P}=&\frac{1}{2|\alpha\beta|\sqrt{a\left(\cosh(v)-1\right)}},\\
\Delta\theta^{(b)}_{QCRB,P}=&\frac{1}{2|\alpha|\sqrt{\frac{a}{2}|\alpha|^2\sinh^2(v)+a|\beta|^2\left(\cosh(v)-1\right)}},\\
\Delta\theta^{(c)}_{QCRB,P}=&\frac{1}{\sqrt{\frac{a}{2}\left(|\alpha|^4+|\beta|^4\right)\sinh^2(v)+2a|\alpha\beta|^2\left(\cosh(v)-1\right)}},
\end{align}
}
In the case of BGCSs, the QFIs take the following analytical form
{\small
\begin{align}
\mathcal{H}^{(a)}_{B}=&4|\xi||\alpha\beta|^{2}\frac{I_{2a}(2|\xi|)}{I_{2a-1}(2|\xi|)},\\
\mathcal{H}^{(b)}_{B}=&\frac{4|\xi|}{I_{2a-1}^{2}(2|\xi|)}\left[|\alpha|^{4}X+|\alpha\beta|^{2}I_{2a-1}(2|\xi|)I_{2a}(2|\xi|)\right],\\
\mathcal{H}^{(c)}_{B}=&\frac{|\xi|}{I_{2a-1}^{2}(2|\xi|)}\left[\left(|\alpha|^4+|\beta|^4\right)X+2|\alpha\beta|^{2}I_{2a-1}(2|\xi|)I_{2a}(2|\xi|) \right],
\end{align}
}
where 
\begin{equation}
X=I_{2a-1}(2|\xi|)\left[|\xi|I_{2a+1}(2|\xi|)+I_{2a}(2|\xi|)\right]-|\xi|I_{2a}^{2}(2|\xi|),
\end{equation}
and
{\small
\begin{align}
\Delta\theta^{(a)}_{QCRB,B}=&\frac{1}{2|\alpha\beta|}\sqrt{\frac{I_{2a-1}(2|\xi|)}{|\xi|I_{2a}(2|\xi|)}},\\
\Delta\theta^{(b)}_{QCRB,B}=&\frac{I_{2a-1}(2|\xi|)}{2|\alpha|\sqrt{|\xi|\left(|\alpha|^{2}X+|\beta|^{2}I_{2a-1}(2|\xi|)I_{2a}(2|\xi|)\right)}},\\
\Delta\theta^{(c)}_{QCRB,B}=&\frac{I_{2a-1}(2|\xi|)}{\sqrt{|\xi|\left[\left(|\alpha|^4+|\beta|^4\right)X+2|\alpha\beta|^{2}I_{2a-1}(2|\xi|)I_{2a}(2|\xi|) \right]}}.
\end{align}
}
In the balanced scenario (i.e., $|\alpha|=|\beta|=1/\sqrt{2}$), the QFIs associated with PCSs takes the form
\begin{align}
\mathcal{H}^{(a)}_{P}=&a\left(\cosh(v)-1\right),\\
\mathcal{H}^{(b)}_{P}=&a\left(\frac{1}{2}\sinh^2(v)+\cosh(v)-1\right),\\
\mathcal{H}^{(c)}_{P}=&\frac{1}{4}a\sinh^2(v)+\frac{a}{2}\left(\cosh(v)-1\right).
\end{align}
The corresponding QCRBs are given by
\begin{align}
\Delta\theta^{(a)}_{QCRB,P}=&\frac{1}{\sqrt{a\left(\cosh(v)-1\right)}},\\
\Delta\theta^{(b)}_{QCRB,P}=&\frac{1}{\sqrt{a\left(\frac{1}{2}\sinh^2(v)+\cosh(v)-1\right)}},
\end{align}
and
\begin{equation}
    \Delta\theta^{(c)}_{QCRB,P}=\frac{1}{\sqrt{\frac{a}{4}\sinh^2(v)+\frac{a}{2}(\cosh(v)-1)}}.
\end{equation}
Similarly for the BGCSs, we can write the analytical expressions as
\begin{align}
\mathcal{H}^{(a)}_{B}=&|\xi|\frac{I_{2a}(2|\xi|)}{I_{2a-1}(2|\xi|)},\\
\mathcal{H}^{(b)}_{B}=&\frac{|\xi|}{I_{2a-1}^{2}(2|\xi|)}\left[X+I_{2a-1}(2|\xi|)I_{2a}(2|\xi|)\right],\\
\mathcal{H}^{(c)}_{B}=&\frac{|\xi|}{2I_{2a-1}^{2}(2|\xi|)}\left[X+I_{2a-1}(2|\xi|)I_{2a}(2|\xi|) \right].
\end{align}
and the corresponding QCRBs take the form
\begin{align}
\Delta\theta^{(a)}_{QCRB,B}=&\sqrt{\frac{I_{2a-1}(2|\xi|)}{|\xi|I_{2a}(2|\xi|)}},\\
\Delta\theta^{(b)}_{QCRB,B}=&\frac{I_{2a-1}(2|\xi|)}{\sqrt{|\xi|\left[X+I_{2a-1}(2|\xi|)I_{2a}(2|\xi|)\right]}},
\end{align}
and
\begin{equation}
    \Delta\theta^{(c)}_{QCRB,B}=\frac{\sqrt{2}I_{2a-1}(2|\xi|)}{\sqrt{|\xi|\left[X+I_{2a-1}(2|\xi|)I_{2a}(2|\xi|) \right]}}.
\end{equation}
\begin{widetext}
	
	\begin{figure}[ht] 
	\begin{minipage}[b]{.50\linewidth}
		\centering
		
		\includegraphics[scale=0.45]{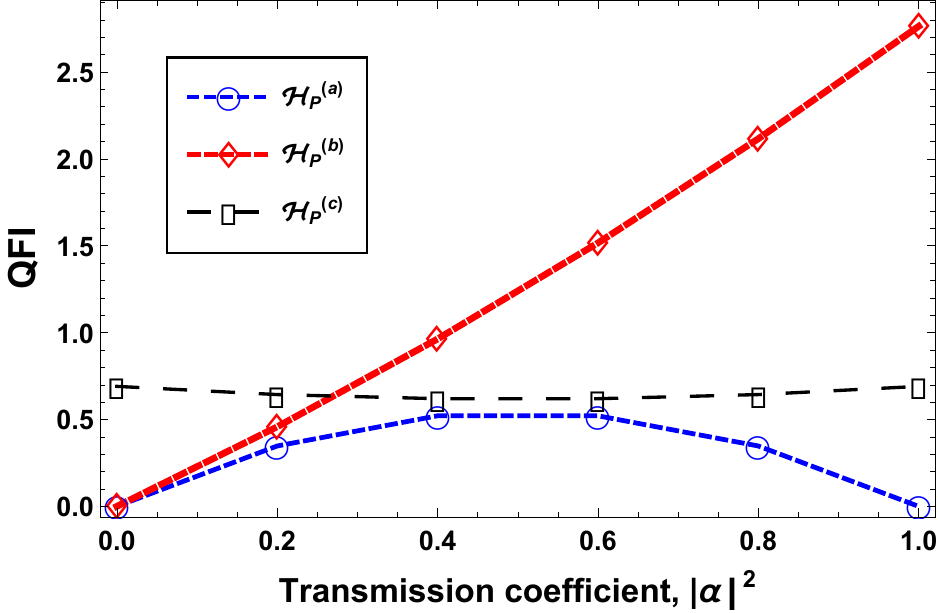} 
  \vfill $\left(a\right)$
	\end{minipage} \quad
	\begin{minipage}[b]{.45\linewidth}
		\centering
		\includegraphics[scale=0.45]{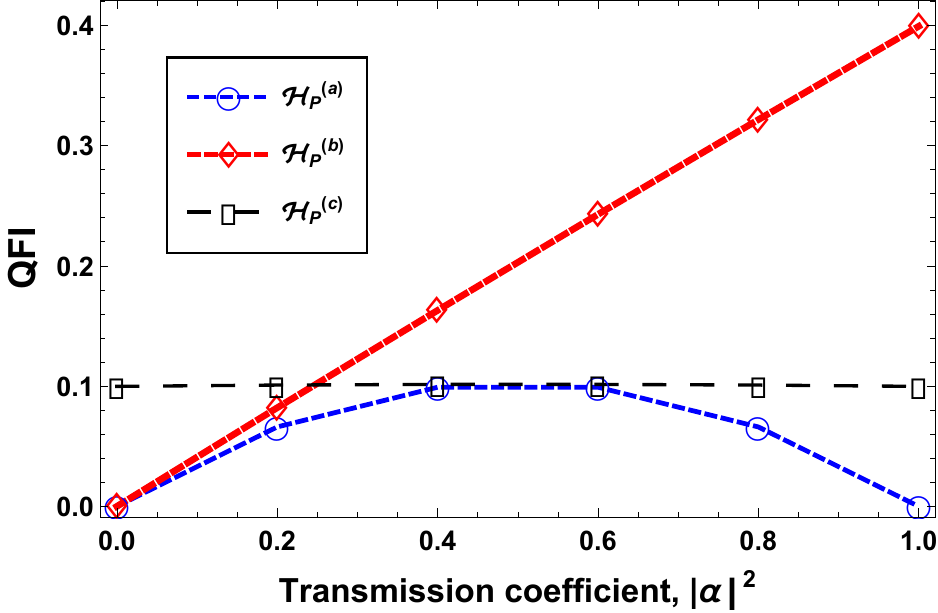} 
  \vfill $\left(b\right)$
	\end{minipage}
	\caption{The dynamics of the three quantum Fisher information metrics, denoted as $\mathcal{H}^{(a)}$, $\mathcal{H}^{(b)}$, and $\mathcal{H}^{(c)}$, with respect to the transmission coefficient $|\alpha|^2$ of the first beam splitter for a single input state; Panel ($a$) for the Perelomov coherent input state with parameters $a=1$ and $v=1$, and panel ($b$) for the Barut-Girardello coherent input state with parameters $a=1$ and $v=1$. The graphical representation distinguishes $\mathcal{H}^{(a)}$, $\mathcal{H}^{(b)}$ and $\mathcal{H}^{(c)}$: the blue line for $\mathcal{H}^{(a)}$, the red line for $\mathcal{H}^{(b)}$, and the long black line for $\mathcal{H}^{(c)}$.} \label{Fig2}
	\end{figure}	
 
\end{widetext}
For the input coherent state $|\psi_{\text{in}}\rangle$, the average number of photons is given by
\begin{equation}
    \langle N \rangle = \langle \psi_{\text{in}} | \hat{b}_{1}^\dagger \hat{b}_{1}| \psi_{\text{in}} \rangle = \langle \hat{g}_1 \rangle.
\end{equation}
In interferometry, it is well known that the minimum phase measurement error of a standard MZI using coherent light is lower bounded by the standard quantum limit, given by $\Delta \phi_{\text{SQL}} \geq 1/\sqrt{\langle N \rangle}$. The average photon number for the PCS is
\begin{equation}
   \langle N \rangle_P=a\left(\cosh(v)-1\right),
\end{equation}
while for the BGCS, it is
\begin{equation}
   \langle N \rangle_B=|\xi|\frac{I_{2a}(2|\xi|)}{I_{2a-1}(2|\xi|)}.
\end{equation}

In Figure (\ref{Fig2}), we illustrate the dynamic behavior of the three QFIs metrics considered, $\mathcal{H}^{(a)}$, $\mathcal{H}^{(b)}$, and $\mathcal{H}^{(c)}$, with respect to the transmission coefficient $|\alpha|^2$ of the first beam splitter for both the PCS and the BGCS with parameters $a=1$ and $v=1$. In Fig.\ref{Fig2}($a$), focusing on the scenario where the single input state is PCS, the plots show that the QFI $\mathcal{H}^{(c)}_{P}$ remains nearly constant for different values of the transmission coefficient $\alpha$, while the QFI $\mathcal{H}^{(b)}_{P}$ exhibits a linear variation, increasing steadily from $0$ for $|\alpha|=0$ to $2a\sinh^2(v)$ for $|\alpha|=1$. Moreover, the two-parameter QFI, represented by $\mathcal{H}^{(a)}_{P}$, reaches its maximum value in the balanced case, i.e., when $|\alpha|=|\beta| = 1/\sqrt{2}$. In the extreme case where $|\alpha|$ equals $0$ or $1$, the QFI reaches its minimum value, $\mathcal{H}^{(a)}_{P}=0$. Moving to Fig.\ref{Fig2}($b$), the QFI is plotted for the single input state BGCS. Here, $\mathcal{H}^{(c)}$ remains constant, while $\mathcal{H}^{(b)}_{P}$ shows a linear progression, consistently ranging from its minimum for $|\alpha|=0$ to its maximum for $|\alpha|=1$, which corresponds to $4|\xi|/I_{2a-1}^{2}(2|\xi|)$. Furthermore, for the two-parameter QFI, $\mathcal{H}^{(a)}{P}$ reaches its maximum value when $|\alpha|=|\beta| = 1/\sqrt{2}$. The QFI vanishes in the extreme cases where $|\alpha|$ equals $0$ or $1$. More importantly, comparing Fig.\ref{Fig2}($a$) with Fig.\ref{Fig2}($b$), we observe that the values of the phase estimate for PCSs are larger than those for BGCSs in all the scenarios considered.

\section{Phase sensitivity in a Mach-Zehnder interferometer}\label{Sec5}
We will now proceed to close the MZI using BS2, which is characterized by its transmission (reflection) coefficient $\alpha'$ ($\beta'$). We will then examine the performance of three realistic detection schemes: namely, the intensity difference, the single-mode intensity, and the balanced homodyne detection. To delve deeper into the analysis, we explore the quantum parameter estimation problem. In this context, we consider an experimentally accessible Hermitian operator, denoted by $\hat{S}$, that depends on a parameter $\theta$. In our specific scenario, $\theta$ corresponds to the phase shift in a MZI and may or may not be observable. Its average is given by
\begin{equation}
    \langle\hat{S}(\theta)\rangle=\langle\psi|\hat{S}(\theta)|\psi\rangle,
\end{equation}
where $|\psi\rangle$ represents the wave function of the system. When a small variation $\delta\theta$ is applied to the parameter $\theta$, it induces a change described as
\begin{equation}
    \langle\hat{S}(\theta+\delta\theta)\rangle \approx \langle\hat{S}(\theta)\rangle+\frac{\partial\langle\hat{S}(\theta)\rangle}{\partial\theta}\delta\theta.
\end{equation}
The experimental detectability of the difference between $ \langle\hat{S}(\theta+\delta\theta)\rangle$ and $\langle\hat{S}(\theta)\rangle$ depends on satisfying the condition
\begin{equation} \label{Condition}
     \langle\hat{S}(\theta+\delta\theta)\rangle - \langle\hat{S}(\theta)\rangle  \geq \Delta\hat{S}(\theta),
\end{equation}
where $\Delta\hat{S}$ is the standard deviation of $\hat{S}$ and is defined as the square root of the variance $\Delta^{2}\hat{S}$, which is expressed as $\Delta\hat{S}=\sqrt{\langle\hat{S}^{2}\rangle - \langle\hat{S}\rangle^{2}}$.
If the inequality (\ref{Condition}) is saturated by the value of $\delta\theta$, then this variation $\delta\theta$ is called the sensitivity, denoted by $\Delta\theta$ \cite{dAriano1994}
\begin{equation}
    \Delta\theta=\frac{\Delta\hat{S}}{|\frac{\partial}{\partial\theta}\langle\hat{S}\rangle|}.
\end{equation}
In the following, $\theta$ represents the total phase shift inside the interferometer, which is divided into two parts: the first part, denoted as $\theta_{i}$, represents the quantity we want to measure, and the second part is $\theta_{exp}$, which is experimentally controllable. We express this relationship as $\theta=\theta_{i}+\theta_{exp}$. In interferometry, the condition $|\theta_{i}|\ll|\theta|$ is crucial because it indicates that the unknown phase shift $\theta_{i}$ has a limited effect on the total phase shift $\theta$. Therefore, the experimenter must adjust $\theta_{exp}$ to approach the optimal phase shift, denoted as $\theta_{opt}$, to achieve the best performance. In the above description, we will focus on the phase sensitivity for each of the considered detection schemes, i.e., intensity difference detection, single-mode intensity detection, and balanced homodyne detection.

\subsection{Intensity difference detection scheme}
In the intensity difference detection scheme, sensitive only to the difference between phase shifts $\theta_{1}$ and $\theta_{2}$, we compute the disparity in the output photocurrents, labeled as $\hat{G}_{dif}$, specifically those detected at ${\rm D}_{4}$ and ${\rm D}_{5}$, as illustrated in Fig.(\ref{Fig1}). Thus, the output operator $\hat{G}_{dif}$ is defined as
\begin{equation}
    \hat{G}_{dif}=\hat{b}^{\dagger}_{4}\hat{b}_{4}-\hat{b}^{\dagger}_{5}\hat{b}_{5}.
\end{equation}
To express the operator $\hat{G}_{dif}$ in terms of the input field operators, we need the field operator transformations,
\begin{equation} \label{transformations2}
    \hat{b}_{4}=\alpha'\hat{b}_{2}+\beta'\hat{b}_{3},\quad \hat{b}_{5}=\beta'\hat{b}_{2}+\alpha'\hat{b}_{3},
\end{equation}
where $\alpha'$ ($\beta'$) represents the transmission (reflection) coefficients of the second beam splitter (BS2). Using the field operator transformations (\ref{transformations1}), we obtain
\begin{align} \label{transformations22}
    &\hat{b}_{4}=[(\alpha\alpha'e^{-i\theta_{2}}+\beta\beta'e^{-i\theta_{1}})\hat{b}_{0}+(\alpha\beta'e^{-i\theta_{1}}+\beta\alpha'e^{-i\theta_{2}})\hat{b}_{1}],\notag\\
    &\hat{b}_{5}=[(\alpha\beta'e^{-i\theta_{2}}+\beta\alpha'e^{-i\theta_{1}})\hat{b}_{0}+(\alpha\alpha'e^{-i\theta_{1}}+\beta\beta'e^{-i\theta_{2}})\hat{b}_{1}],
\end{align}
where $\theta=\theta_{1}-\theta_{2}$. Substituting the field operator transformations into the definition of $\hat{G}_{dif}$ yields
{\small
\begin{align}
    \hat{G}_{dif}&=\left[(2|\alpha|^{2}-1)(2|\alpha'|^{2}-1)-4|\alpha\alpha'\beta\beta'|\cos\theta\right](\hat{g}_{0}-\hat{g}_{1})\notag\\ \nonumber
    &+4\mathfrak{R}\left\{\left((|\beta|^{2}e^{-i\theta}-|\alpha|^{2}e^{i\theta})\alpha'^{\ast}\beta'+\alpha^{\ast}\beta(1-2|\alpha'|^{2})\right)\hat{b}_{0}\hat{b}_{1}^{\dagger}\right\}.
\end{align}}
To quantify the phase sensitivity here, we can define it as
\begin{equation}
    \Delta\theta_{dif}=\frac{\Delta\hat{G}_{dif}}{|\frac{\partial}{\partial\theta}\langle\hat{G}_{dif}\rangle|},
\end{equation}
where the derivative of $\langle\hat{G}_{dif}\rangle$ with respect to $\theta$ is given by
\begin{align}
    \frac{\partial}{\partial\theta}\langle\hat{G}_{dif}\rangle&=4|\alpha\alpha'\beta\beta'|\sin\theta(\langle\hat{g}_{0}\rangle-\langle\hat{g}_{1}\rangle)\notag\\
    &+4|\alpha'\beta'|\mathfrak{R}\{(|\beta|^{2}e^{-i\theta}+|\alpha|^{2}e^{i\theta})\langle\hat{b}_{0}\rangle\langle\hat{b}^{\dagger}_{1}\rangle\}.
\end{align}
Thus, the variance of the operator $\hat{G}_{dif}$ can be derived as the following form
\begin{align}
    \Delta^{2}\hat{G}_{dif}=&\delta_{A}^{2}(\Delta^{2}\hat{g}_{0}+\Delta^{2}\hat{g}_{1})+|\delta_{B}|^{2}(\langle\hat{g}_{0}\rangle+\langle\hat{g}_{1}\rangle)\notag\\ \nonumber &+2|\delta_{B}|^{2}\left(\langle\hat{g}_{0}\rangle\langle\hat{g}_{1}\rangle-|\langle\hat{b}_{0}\rangle|^{2}||\langle\hat{b}_{1}\rangle|^{2}\right)\\ \nonumber
    &+2\mathfrak{R}\left\{\delta_{B}^{2}(\langle\hat{b}_{0}^{2}\rangle\langle(\hat{b}_{1}^{\dagger})^{2}\rangle)-\langle\hat{b}_{0}\rangle^{2}\langle\hat{b}_{1}^{\dagger}\rangle^{2}\right\}\\ \nonumber
    &+4\delta_{A}\mathfrak{R}\left\{\delta_{B}\left((\langle\hat{g}_{0}\hat{b}_{0}\rangle-\langle\hat{g}_{0}\rangle\langle\hat{b}_{0}\rangle)\langle\hat{b}_{1}^{\dagger}\rangle\right.\right.\\
    &\qquad \left.\left.-\langle\hat{b}_{0}\rangle(\langle\hat{b}_{1}^{\dagger}\hat{g}_{1}\rangle-\langle\hat{g}_{1}\rangle\langle\hat{b}_{1}^{\dagger}\rangle)\right)\right\}, 
\end{align}
where
\begin{align} \label{Ad and Cd}
\delta_{A}=&1-2\left(|\alpha||\beta'|+|\beta||\alpha'|\right)^{2}+4|\alpha\beta||\alpha'\beta'|(1-\cos\theta),\\
\delta_{B}=&2i\left(|\alpha\beta|(1-2|\alpha'|^{2})+(1-2|\alpha|^{2})|\alpha'\beta'|\cos\theta\right)\\ \nonumber
&+2|\alpha'\beta'|\sin\theta,
\end{align}
satisfy the following condition
\begin{equation}
    \delta_{A}^{2}+|\delta_{B}|^{2}=1.
\end{equation}
In this detection scheme, the phase sensitivity remains the same for both scenarios ($b$) and ($c$). For simplicity, we will use the notation $\Delta\theta_{dif}$ to collectively represent the phase sensitivity in all detection schemes.

\subsection{Single-mode intensity detection scheme}
In the single-mode intensity detection scheme, we focus on a single photocurrent at output port $4$ (see Fig.(\ref{Fig1})), represented by its associated operator $\hat{g}_{4}=\hat{b}_{4}^{\dagger}\hat{b}_{4}$. The phase sensitivity in this scenario is defined as
\begin{equation}
    \Delta\theta_{sing}=\frac{\Delta\hat{g}_{4}}{|\frac{\partial}{\partial\theta}\langle\hat{g}_{4}\rangle|}.
\end{equation}
From equation (\ref{transformations22}), we can determine the average number of photons with respect to the input field operator as
{\small
\begin{align}
    \langle\hat{g}_{4}\rangle&=\left(|\alpha\alpha'|^{2}+|\beta\beta'|^{2}-2|\alpha\alpha'\beta\beta'|\cos\theta\right)\langle\hat{g}_{0}\rangle\\ \nonumber
    &+\left(|\alpha\beta'|^{2}+|\alpha'\beta|^{2}+2|\alpha\alpha'\beta\beta'|\cos\theta\right)\langle\hat{g}_{1}\rangle\\ \nonumber
    &+2\mathfrak{R}\left\{\left(\alpha^{\ast}\beta(2|\alpha'|^{2}-1)+\alpha'^{\ast}\beta'e^{-i\theta}(|\alpha|^{2}-|\beta|^{2}e^{2i\theta})\right)\langle\hat{b}_{0}^{\dagger}\rangle\langle\hat{b}_{1}\rangle\right\}.
\end{align}
}
Using the above equation, we immediately get
\begin{align}
\frac{\partial\langle\hat{g}_{4}\rangle}{\partial\theta}=&2|\alpha\alpha'\beta\beta'|\sin\theta(\langle\hat{g}_{0}\rangle-\langle\hat{g}_{1}\rangle)\\
&+2||\alpha'^{\ast}\beta'|\mathfrak{R}\left\{\left(|\alpha|^{2}e^{-i\theta}+|\beta|^{2}e^{i\theta}\right)\langle\hat{b}_{0}^{\dagger}\rangle\langle\hat{b}_{1}\rangle\right\}.\nonumber
\end{align}
To find $\Delta^{2}\hat{g}_{4}$, we first calculate the square of the operator $\hat{g}_{4}$. Then we get the final expression for $\Delta^{2}\hat{g}_{4}$ as
\begin{align}\nonumber
\Delta^{2}\hat{g}_{4}&=|\delta_{3}|^{2}\left(\langle\hat{g}_{0}\rangle+\langle\hat{g}_{1}\rangle+2\langle\hat{g}_{0}\rangle\langle\hat{g}_{1}\rangle-2|\langle\hat{b}_{0}\rangle|^{2}|\langle\hat{b}_{1}\rangle|^{2}\right)\\ \nonumber
&+2\delta_{0}\mathfrak{R}\left\{\delta_{3}(\langle\hat{g}_{0}\hat{b}_{0}^{\dagger}\rangle+\langle\hat{b}_{0}^{\dagger}\hat{g}_{0}\rangle-2\langle\hat{g}_{0}\rangle\langle\hat{b}_{0}^{\dagger}\rangle)\langle\hat{b}_{1}\rangle\right\}\\ \nonumber
&+2\delta_{1}\mathfrak{R}\left\{\delta_{3}\langle\hat{b}_{0}^{\dagger}\rangle(\langle\hat{g}_{1}\hat{b}_{1}\rangle+\langle\hat{b}_{1}\hat{g}_{1}\rangle-2\langle\hat{g}_{1}\rangle\langle\hat{b}_{1}\rangle)\right\}\\ \nonumber
&+2|\delta_{3}^{2}|\mathfrak{R}\left\{\langle\hat{b}_{0}^{2}\rangle\langle(\hat{b}_{1}^{\dagger})^{2}\rangle-\langle\hat{b}_{0}\rangle^{2}\langle\hat{b}_{1}^{\dagger}\rangle^{2}\right\}+\delta_{0}^{2}\Delta^{2}\hat{g}_{0}\\ 
&+\delta_{1}^{2}\Delta^{2}\hat{g}_{1}, 
\end{align}
where
\begin{align} \label{A1}
\delta_{0}=&|\alpha\alpha'|^{2}+|\beta\beta'|^{2}-2|\alpha\alpha'\beta\beta'|\cos\theta,\\
\delta_{1}=&|\alpha\beta'|^{2}+|\alpha'\beta|^{2}+2|\alpha\alpha'\beta\beta'|\cos\theta,\\
\delta_{3}=&\alpha^{\ast}\beta(2|\alpha'|^{2}-1)+\alpha'^{\ast}\beta'(|\alpha|^{2}e^{-i\theta}-|\beta|^{2}e^{i\theta}).
\end{align}

\subsection{Balanced homodyne detection scheme}
We now turn to the balanced homodyne detection scheme at output port 4 (see Fig.(\ref{Fig1})). The operator of interest for modeling this detection scheme is given by
\begin{equation}
    \hat{X}_{\theta_{L}}=\mathfrak{R}\left\{e^{-i\theta_{L}}\hat{b}_{4}\right\},
\end{equation}
where $\theta_L$ is the phase of the local coherent source $|\gamma\rangle$, where $|\gamma\rangle = |\gamma|e^{i\theta_L}$ and $\gamma$ is a complex number. In this detection scheme, we define the phase sensitivity as follows
\begin{equation}
    \Delta\theta_{hom}=\frac{\sqrt{\Delta^{2}\hat{X}_{\theta_{L}}}}{|\frac{\partial\langle\hat{X}_{\theta_{L}}\rangle}{\partial\theta}|}.
\end{equation}
Using the field operator transformations (\ref{transformations22}), we arrive at the final expression for $\langle\hat{X}_{\theta_{L}}\rangle$ as
\begin{align}
    \langle\hat{X}_{\theta_{L}}\rangle=&\mathfrak{R}\left\{e^{-i\theta_{L}}\left((\alpha\alpha'e^{-i\theta_{2}}+\beta\beta'e^{-i\theta_{1}})\langle\hat{b}_{0}\rangle \right.\right.\\
    &\left.\left. +(\alpha\beta'e^{-i\theta_{1}}+\beta\alpha'e^{-i\theta_{2}})\langle\hat{b}_{1}\rangle\right)\right\}, \nonumber
\end{align}
and the variance of the above operator is given by
\begin{align}
\Delta^{2}\hat{X}_{\theta_{L}}=&\frac{1}{4}+2\mathfrak{R}\left\{x^{2}\Delta^{2}\hat{b}_{0}+y^{2}\Delta^{2}\hat{b}_{1}\right\}\\
&+2|x|^{2}(\langle\hat{g}_{0}\rangle-|\langle\hat{b}_{0}\rangle|^{2})+2|y|^{2}(\langle\hat{g}_{1}\rangle-|\langle\hat{b}_{1}\rangle|^{2}), \nonumber
\end{align}
where the coefficients $x$ and $y$ are given by
\begin{align} \label{A and B}
x=&\frac{1}{2}e^{-i(\theta_{L}+\theta_{2})}\left(\alpha\alpha'+\beta\beta'e^{-i\theta}\right),\\
y=&\frac{1}{2}e^{-i(\theta_{L}+\theta_{2})}\left(\alpha\beta'e^{-i\theta}+\alpha'\beta\right).&
\end{align}
For scenario ($b$) in Fig.(\ref{Fig1}), where $\theta_{1}=\theta$ and $\theta_{2}=0$, the absolute value of the derivative of $\langle\hat{X}_{\theta_{L}}\rangle$ with respect to $\theta$ is given by
\begin{equation} \label{di}
    |\frac{\partial\langle\hat{X}_{\theta_{L}}\rangle}{\partial\theta}|=|\mathfrak{R}\left\{e^{-i(\theta_{L}+\theta)}\left(\beta\langle\hat{b}_{0}\rangle+\alpha\langle\hat{b}_{1}\rangle\right)\right\}||\beta'|,
\end{equation}
and for scenario ($c$), where $\theta_{1}=-\theta_{2}=\theta/2$, we obtain
\begin{align} \label{dii}
    |\frac{\partial\langle\hat{X}_{\theta_{L}}\rangle}{\partial\theta}|=&\frac{1}{2}|\mathfrak{R}\left\{ie^{-i\theta_{L}}\left(\left(\alpha\alpha'e^{i\theta/2}-\beta\beta'e^{-i\theta/2}\right)\langle\hat{b}_{0}\rangle\right.\right.\notag\\
     &\left.\left. +\left(\beta\alpha'e^{i\theta/2}-\alpha\beta'e^{-i\theta/2}\right)\right)\langle\hat{b}_{1}\rangle\right\}|.
\end{align}

\section{Phase sensitivity with SU(1,1) coherent states in the input}\label{Sec6}
In this section, we compare the phase sensitivities achievable by the three considered detection schemes, as presented in Section IV, for input SU(1,1) coherent states with the QCRBs implied by the various QFIs discussed in Section III. Using the results of the phase sensitivities reported in the previous section, i.e., $\Delta\theta_{dif}$, $\Delta\theta_{sing}$, and $\Delta\theta_{hom}$, and considering the input state (\ref{input state}), it is easy to verify that the phase sensitivities in our input states are as follows:\par
For a intensity difference detection scheme, we get the final analytical expression of the phase sensitivity for the two types of SU(1,1) CSs, as
\begin{equation}
    \Delta\theta_{dif}^{i}=\frac{\Delta_{i}\hat{G}_{dif}}{4|\alpha\alpha'\beta\beta'||\sin\theta \langle\hat{g}_{1}\rangle_{i}|},
\end{equation}
where the subscript $i=P$ or $B$ corresponds to PCS (Eq.\ref{PCS}) and BGCS (Eq.\ref{BGCS}), respectively. Then, from the above expression of the phase sensitivity, we calculate its final analytical expression for the two types of SU(1,1) CSs, as
\begin{equation}
    \Delta\theta_{dif}^{P}=\frac{\sqrt{\frac{1}{2}\delta_{A}^{2}\sinh^{2} v+|\delta_{B}|^{2}(\cosh v-1)}}{4\sqrt{a}|\alpha\alpha'\beta\beta'||(\cosh v-1)\sin\theta|},
\end{equation}
\begin{align}
    \Delta\theta_{dif}^{B}=&\frac{\sqrt{\delta_{A}^{2}|\xi|\left[ I_{2a-1}I_{2a+1}-I_{2a}^{2}\right]+I_{2a-1}I_{2a}}}{4|\alpha\alpha'\beta\beta'||\sin\theta|\sqrt{|\xi|}I_{2a}}, 
\end{align}
where $\delta_{A}$ and $\delta_{B}$ were defined in equation (\ref{Ad and Cd}). To simplify notation, we define $I_{2a} \equiv I_{2a}(2|\xi|)$. Interestingly, for both scenarios ($b$) and ($c$), this detection scheme yields the same phase sensitivity result.\par
For a single-mode intensity detection scheme, we obtain the phase sensitivity in all considered scenarios as
\begin{equation}
    \Delta\theta_{sing}^{i}=\frac{\Delta_{i}\hat{g}_{4}}{2|\alpha\alpha'\beta\beta'||\sin\theta \langle\hat{g}_{1}\rangle_{i}|},
\end{equation}
and the analytical expression of the phase sensitivity for our input SU(1,1) coherent states takes the form
\begin{equation}
    \Delta\theta_{sing}^{P}=\frac{\sqrt{\frac{1}{2}\delta_{1}^{2}\sinh^{2}v+|\delta_{3}|^{2}(\cosh v-1)}}{2\sqrt{a}|\alpha\alpha'\beta\beta'||\sin\theta (\cosh v-1)|},
\end{equation}
{\small
\begin{align}
    \Delta\theta_{sing}^{B}=\frac{\sqrt{\delta_{1}^{2}|\xi|\left[ I_{2a-1}I_{2a+1}-I_{2a}^{2}\right]+(\delta_{1}^{2}+|\delta_{3}|^{2})I_{2a-1}I_{2a}}}{2\sqrt{|\xi|}|\alpha\alpha'\beta\beta'||\sin\theta|I_{2a}},
\end{align}
}
where $\delta_{1}$ and $\delta_{3}$ were defined in equation (\ref{A1}). Finally, for a  balanced homodyne detection scheme, we have
\begin{equation}
\Delta^{2}_{i}\hat{X}_{\theta_{L}}=\frac{1}{4}+2\mathfrak{R}\left\{y^{2}\Delta^{2}_{i}\hat{b}_{1}\right\}+2|y|^{2}(\langle\hat{g}_{1}\rangle_{i}-|\langle\hat{b}_{1}\rangle_{i}|^{2}).
\end{equation}
In the case of scenario ($b$), where $\theta_{1}=\theta$ and $\theta_{2}=0$, and assuming $\theta_{L}=\varphi$, the variance of the operator $\hat{X}_{\theta_{L}}$ is given by

\begin{align}
    \Delta_{i}^{2}\hat{X}_{\theta_{L}}=&\frac{1}{4}-\frac{1}{2}\left(|\alpha\beta'|^{2}\cos2\theta+|\alpha'\beta|^{2}+2|\alpha\alpha'\beta\beta'|\cos\theta\right)\mu_{i}\notag\\
    &+\frac{1}{2}\left(|\alpha\beta'|^{2}+|\alpha'\beta|^{2}+2|\alpha\alpha'\beta\beta'|\cos\theta\right)\left(	\bar{g}_{i}-|\nu_{i}|^{2}\right),
\end{align}
where
\begin{widetext}
{\small
\begin{align} 
&\mu_{P}=\frac{\cosh^{-4a}(v/2)}{\tanh^{2}(v/2)}\left[\sum_{g=2}^{\infty}\frac{\sqrt{\Gamma(g+2a)\Gamma(g+2a-2)}}{\Gamma(2a)(g-2)!}\tanh^{2g}(v/2)-\frac{1}{\cosh^{4a}(v/2)}\left(\sum_{g=1}^{\infty}\frac{\sqrt{\Gamma(g+2a)\Gamma(g+2a-1)}}{\Gamma(2a)(g-1)!}\tanh^{2g}(v/2)\right)^{2}\right],\\
&\mu_{B}=\frac{\tanh^{2a-3}(v/2)}{I_{2a-1}}\left[\sum_{g=2}^{\infty}\frac{\tanh^{2g}(v/2)}{(g-2)!\sqrt{\Gamma(g+2a)\Gamma(g+2a-2)}}-\frac{\tanh^{2a-1}(v/2)}{I_{2a-1}}\left(\sum_{g=1}^{\infty}\frac{\tanh^{2g}(v/2)}{(g-1)!\sqrt{\Gamma(g+2a)\Gamma(g+2a-1)}}\right)^{2}\right],\\
&\nu_{P}=\left(1-\tanh^{2}(v/2)\right)^{2a}\sum_{g=1}^{\infty}\frac{\sqrt{\Gamma(g+2a)\Gamma(g+2a-1)}}{\Gamma(2a)(g-1)!}\tanh^{(2g-1)}(v/2),\\
&\nu_{B}=\frac{\tanh^{2(a-2)}(v/2)}{I_{2a-1}}\sum_{g=1}^{\infty}\frac{\tanh^{2g}(v/2)}{(g-1)!\sqrt{\Gamma(g+2a)\Gamma(g+2a-1)}},\hspace{1cm}\bar{g}_{P}=a(\cosh v-1), \quad \mbox{and} \quad \bar{g}_{B}=\frac{I_{2a}}{I_{2a-1}}\tanh(v/2).
\end{align}
}
\end{widetext}
From equation (\ref{di}), we get
\begin{equation}
    |\frac{\partial\langle\hat{X}_{\theta_{L}}\rangle}{\partial\theta}|=\frac{1}{|\tan(\frac{v}{2})|}|\alpha\beta'||\cos(\theta) \nu_{i}|.
\end{equation}
Based on these results, the phase sensitivity is given by
\begin{align}
    \Delta\theta_{hom}^{(b)}=&|\tan(\frac{v}{2})|\frac{\sqrt{ \Delta_{i}^{2}\hat{X}_{\theta_{L}}}}{|\alpha\beta'||\cos(\theta) \nu_{i}|}.\label{eq116}
\end{align}
In the case of scenario ($c$), where $\theta_{1}=-\theta_{2}=\theta/2$, the above variance is given by
\begin{align}
    \Delta_{i}^{2}\hat{X}_{\theta_{L}}&=\frac{1}{4}-\frac{1}{2}\left((|\alpha\beta'|^{2}+|\alpha'\beta|^{2})\cos\theta+2|\alpha\alpha'\beta\beta'|\right)\mu_{i}\notag\\
    &+\frac{1}{2}\left(|\alpha\beta'|^{2}+|\alpha'\beta|^{2}+2|\alpha\alpha'\beta\beta'|\cos\theta\right)\left(	\bar{g}_{i}-|\nu_{i}|^{2}\right).
\end{align}
Based on the same reasoning, from equation (\ref{dii}) we obtain
\begin{equation}
    |\frac{\partial\langle\hat{X}_{\theta_{L}}\rangle}{\partial\theta}|=\frac{1}{2|\tan(\frac{v}{2})|}||\alpha'\beta|-|\alpha\beta'|||\cos(\frac{\theta}{2}) \nu_{i}|.
\end{equation}
From this, we can calculate the phase sensitivity in this last scenario as
\begin{align}
    \Delta\theta_{hom}^{(c)}=&|2\tan(\frac{v}{2})|\frac{\sqrt{ \Delta_{i}^{2}\hat{X}_{\theta_{L}}}}{||\alpha'\beta|-|\alpha\beta'|||\cos(\frac{\theta}{2}) \nu_{i}|}.\label{eq119}
\end{align}

\begin{widetext}
	
	\begin{figure}[ht] 
	\begin{minipage}[b]{.50\linewidth}
		\centering
		\includegraphics[scale=0.27]{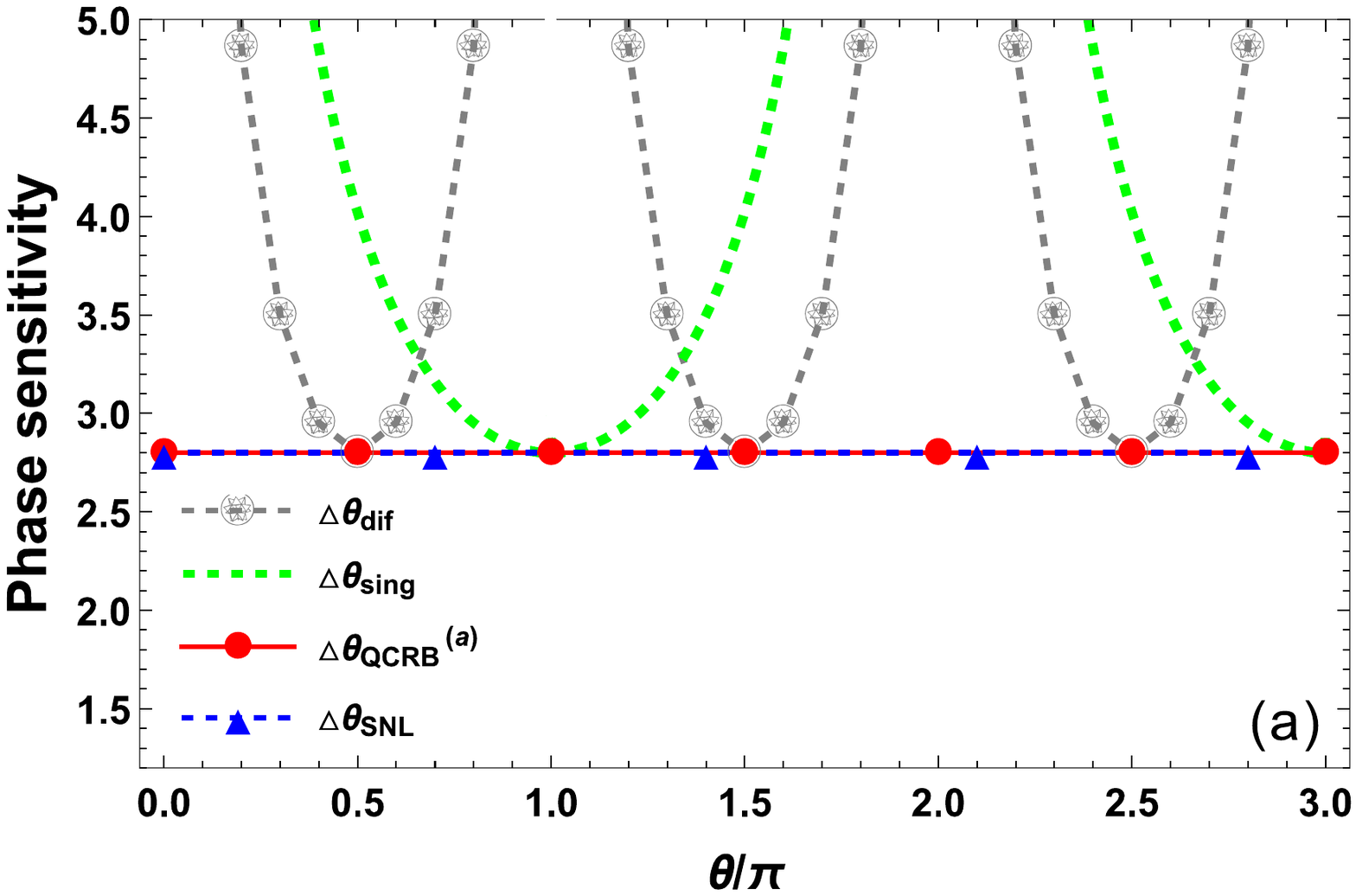} \includegraphics[scale=0.27]{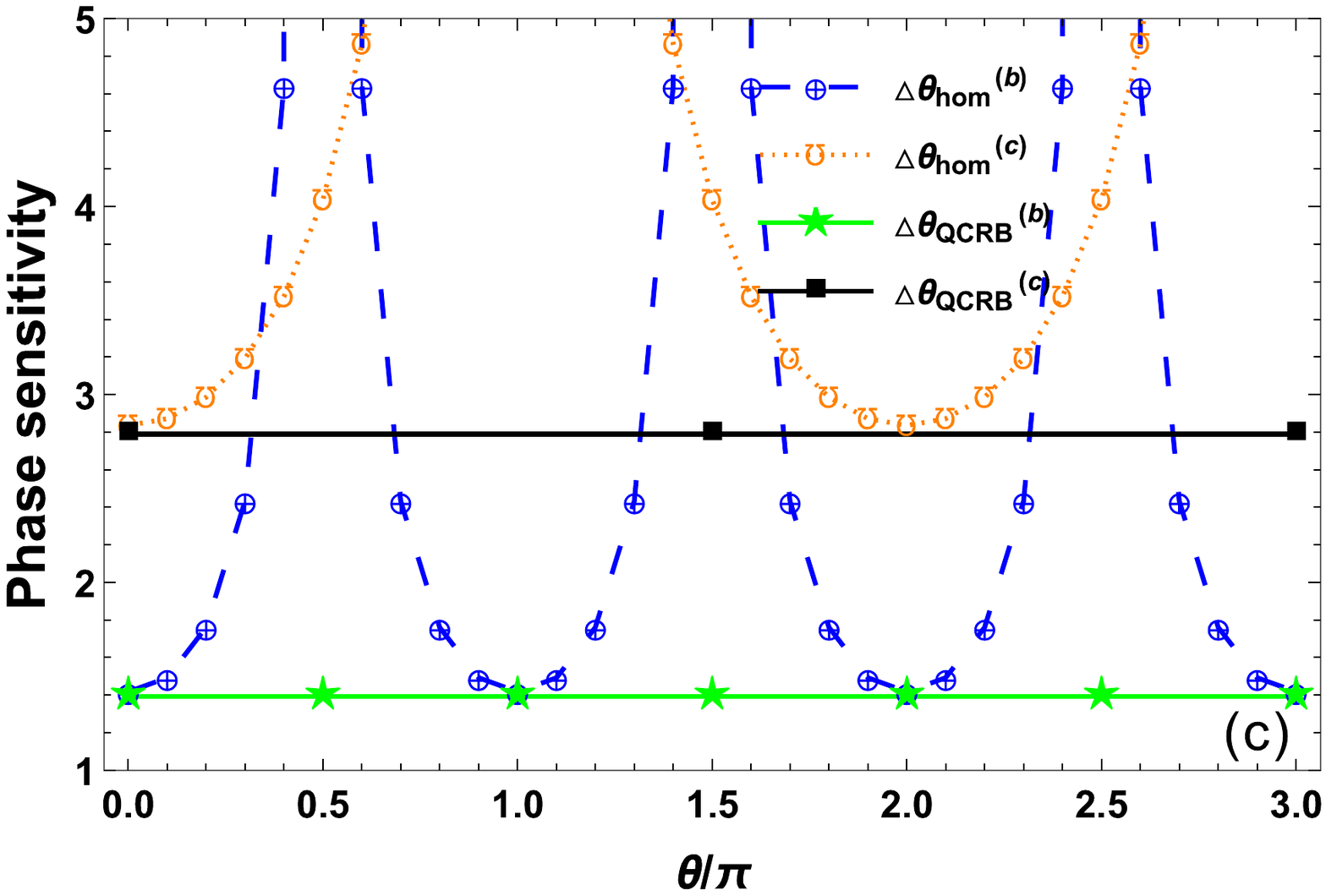} 
	\end{minipage} \quad
	\begin{minipage}[b]{.45\linewidth}
		\centering
		\includegraphics[scale=0.27]{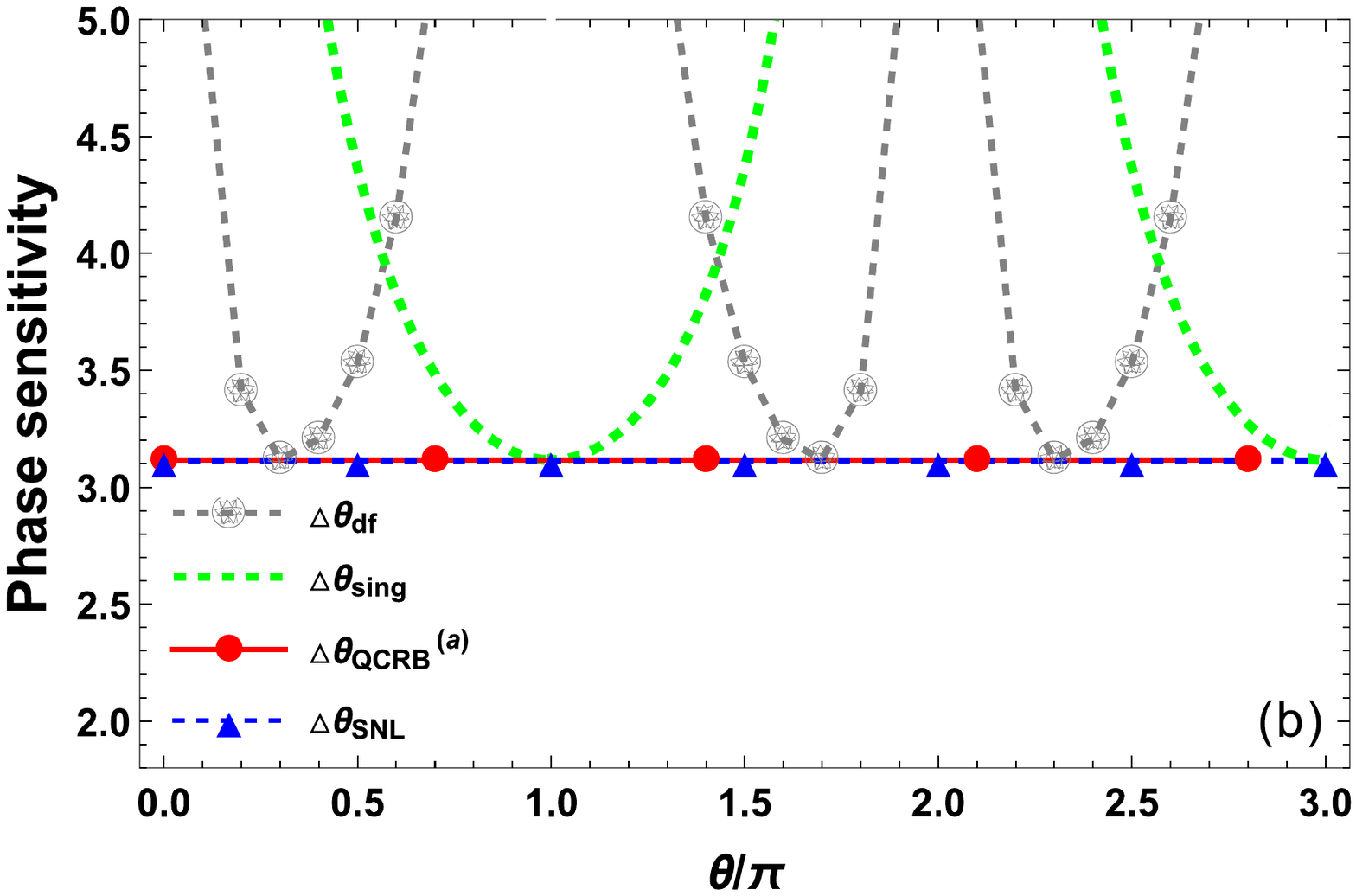} \includegraphics[scale=0.27]{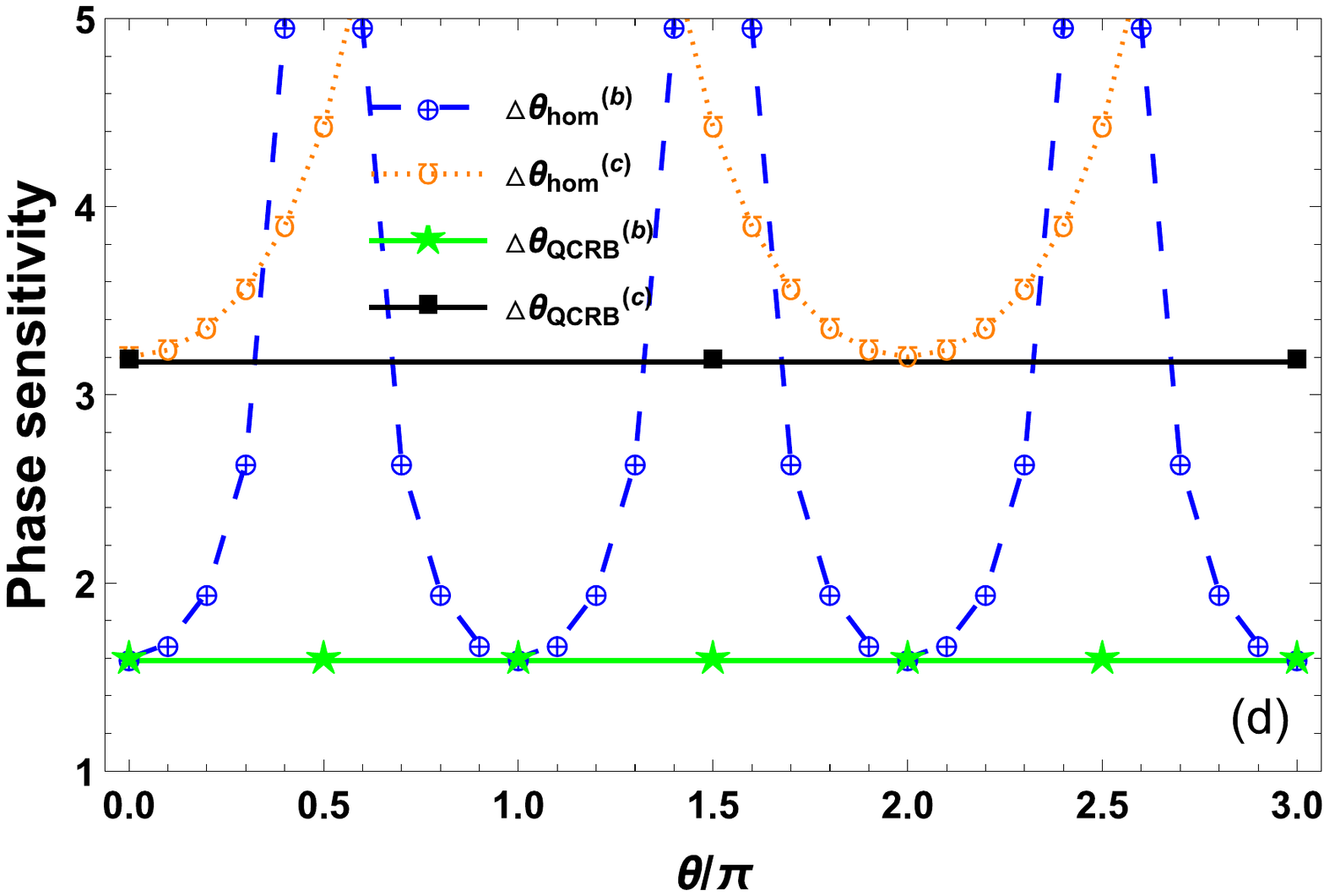} 
	\end{minipage}
	\caption{The phase sensitivity $\Delta\theta$ with respect to the phase shift in the intensity difference detection, single-mode detection, and balanced homodyne detection schemes, denoted as $\Delta\theta_{dif}$, $\Delta\theta_{sing}$, $\Delta\theta_{hom}^{(b)}$, and $\Delta\theta_{hom}^{(c)}$ respectively, together with the three QCRBs $\Delta\theta_{QCRB}^{(a)}$, $\Delta\theta_{QCRB}^{(b)}$, and $\Delta\theta_{QCRB}^{(c)}$. In panels ($a$) and ($c$), for the Perelomov coherent input state with parameters $a=1$ and $v=0.5$, and in panels ($b$) and ($d$), for the Barut-Girardello coherent input state with parameters $a=1$ and $v=1$, the graphical representation includes a gray curve for $\Delta\theta_{dif}$, the green curve for $\Delta\theta_{sing}$, the blue curve for $\Delta\theta_{hom}^{(b)}$, the orange curve for $\Delta\theta_{hom}^{(c)}$, the red line for $\Delta\theta_{QCRB}^{(a)}$, the green line for $\Delta\theta_{QCRB}^{(b)}$, and the black line for $\Delta\theta_{QCRB}^{(c)}$. Additionally, the SNL is represented in panels (a) and (b).} \label{Fig3}
	\end{figure}	
\end{widetext}	

\begin{figure}[ht]
		\includegraphics[scale=0.4]{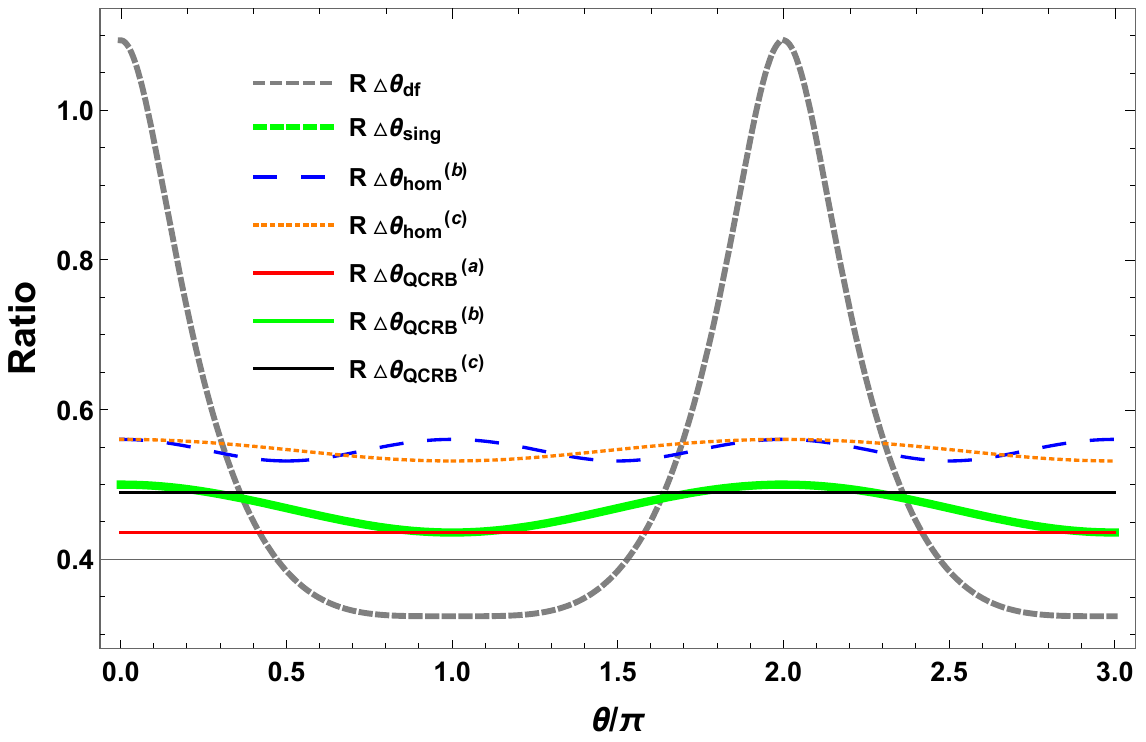} 
	\caption{The performance ratio $R=\Delta\theta_{i}/\Delta\theta_{j}$, with $i$ representing the Perelomov coherent input state and $j$ representing the Barut-Girardello coherent input state, of optimal phase estimation in different detection schemes for the two types of SU(1,1) CSs with $v=1$} \label{Fig4}
\end{figure}
 
Fig.(\ref{Fig3}) presents the variation of the four phase sensitivities as a function of the MZI's phase shift for both the Perelomov and Barut-Girardello coherent input states with parameters $a=1$ and $v=1$. The figure shows these phase sensitivities for the three detection schemes discussed previously, alongside the corresponding three QCRBs.\par

We plot the intensity difference detection and single-mode detection schemes for the case of a balanced beam splitter, i.e., $\alpha=1/\sqrt{2}$ and $\beta=i/\sqrt{2}$, considering its optimal setup. For the balanced homodyne detection scheme, we have chosen the transmission coefficients $\alpha=1$ for the first beam splitter and $\alpha'=0$ for the second beam splitter to achieve optimal phase sensitivity of the interferometer. This configuration minimizes the quantities described in equations (\ref{eq116}) and (\ref{eq119}). The gray curve and the green curve in Fig.\ref{Fig3}(a) and Fig.\ref{Fig3}(b) correspond to the intensity difference detection scheme and the single-mode detection scheme, respectively. As can be seen in this figure, both curves have an optimum that achieves the QCRB, which is implied by the two-parameter QFI. The blue and orange lines are for two balanced homodyne detection schemes. As observed in this figure, these phase sensitivities reach the QCRBs. For the chosen values, the phase sensitivity $\Delta\theta_{hom}^{(b)}$ shows significantly better performance than the phase sensitivity $\Delta\theta_{dif}$. This suggests an advantage to using an external phase reference for both the Perelomov coherent input state and the Barut-Girardello coherent input state. The input state, including the local oscillator, can be expressed as 
\begin{equation}
|\psi\rangle=|\psi_{in}\rangle\otimes|\gamma\rangle=|\xi_{i},a\rangle\otimes||\gamma|e^{i\theta_L}\rangle.
\end{equation}
Our interferometer consists of two arms: one involving input port 1, passing through the first beam splitter with total transmission, phase shift $\theta_1$, the second beam splitter with total reflection, and reaching the beam splitter (BSL). The other arm represents the local oscillator fed into the balanced beam splitter of the homodyne setup.\par

Recent studies have explored various methods to enhance phase sensitivity in MZI, especially by using nonclassical input states and advanced detection schemes. For instance, squeezing-enhanced linear interferometers have been shown to improve sensitivity by combining outputs from both interferometer arms, even under detection losses, as discussed in \cite{Shukla2020}. Additionally, the use of Schrödinger’s cat-like states and Fock states as inputs in MZIs has led to better phase sensitivity under certain conditions, with a two-channel detection scheme proving optimal, as demonstrated in \cite{Shukla2021}. Furthermore, the application of squeezed Kerr states (SKS) has shown promise in achieving superior phase sensitivity in both lossless and lossy conditions, as highlighted in \cite{Yadav2024}. Investigations into quantum LiDAR systems using multi-photon states have also revealed improvements in phase sensitivity, supporting the idea that nonclassical resources such as these can enhance interferometric precision, as seen in \cite{Sharma2024}. Moreover, Shukla et al.\cite{Shukla2023} investigate the use of cat-like states in MZIs, emphasizing their potential for quantum sensing applications. These studies collectively underscore the critical role of nonclassical input states and optimized detection schemes in surpassing classical phase sensitivity limits and advancing quantum sensing technologies.\par

When considering phase estimation as a problem involving two parameters and assuming a vacuum state at input port $0$, equation (\ref{H(a)}) yields the quantum Fisher information $\mathcal{H}^{(a)}=4|\alpha\beta|^2 \langle \hat{g}_1 \rangle$. As the maximum value of $4|\alpha\beta|^2$ is $1$, sub-SNL performance is generally unattainable. This aligns with the claim made in reference \cite{Takeoka2017}, which states that when both arms of the MZI experience different unknown phase shifts and one input port is vacuum, achieving phase sensitivity beyond the SNL is impossible, regardless of the input in the other port or the detection scheme. This scenario is particularly relevant to applications like gravitational wave detection, where maintaining optimal sensitivity is critical for detecting faint signals over substantial noise. However, an exception arises when the two unknown phase shifts are correlated, such that $\theta_1=\theta/2$ and $\theta_2 =-\theta/2$. In this correlated phase-shift scenario, the relevant Fisher information is $\mathcal{H}^{(c)}$ instead of $\mathcal{H}^{(a)}$. For the vacuum state at input port $0$, equation (\ref{H(c)}) provides the appropriate Fisher information for this case. Notably, sub-SNL sensitivity becomes achievable if $\Delta^{2}\hat{g}_{1} > \langle\hat{g}_1\rangle$. This finding underscores the importance of exploiting correlated phase relationships to overcome limitations imposed by standard quantum bounds, providing pathways for innovative applications in precision metrology and quantum sensing.\par

In our setup for homodyne detection using SU(1,1) coherent states, it is important to keep the phase of the signal and the local oscillator synchronized. This is done by using a strong local oscillator that is locked in phase with the signal. In some cases, like when squeezing is involved, keeping the phase between the signal and local oscillator consistent over time is crucial. To ensure this, we use the same laser to create both the local oscillator and to interact with the quantum system, such as an atom, a quantum oscillator, or a nonlinear medium. This setup helps keep the signal and local oscillator correlated, meaning their phases are related in a predictable way.\par

To compare the performance of optimal phase estimation in different detection schemes for the two types of SU(1,1) CSs, we use a technique where we introduce the ratio between the phase sensitivities of these two states in different detection schemes as $R=\Delta\theta_{i}/\Delta\theta_{j}$, with $i$ representing the Perelomov coherent input state and $j$ representing the Barut-Girardello coherent input state. As a result, when the ratio $R<1$, the error limit of the phase sensitivity for the Perelomov coherent input state in different detection schemes is smaller and offers an advantage over that of the Barut-Girardello coherent input state. As shown in Fig.(\ref{Fig4}), we observe that the values of the phase sensitivities for the two states satisfy the inequality $\Delta\theta_{P}<\Delta\theta_{B}$, indicating that the performance of the phase sensitivities for the Perelomov coherent input state is better and would provide a more precise result than that of the Barut-Girardello coherent input state.

\section{CONCLUSION}\label{Sec7}
Optimizing the sensitivity of a MZI requires careful consideration of both the input state and the detection scheme. QFI serves as a valuable tool to identify the optimal operating points that achieve the highest possible sensitivity. This paper presents theoretical calculations of QCRBs for both two-parameter and single-parameter estimation in quantum interferometry, considering two input scenarios. We explore the performance of Perelomov and Barut-Girardello coherent input states within the SU(1,1) Lie algebra. We investigate their phase sensitivity across various detection schemes, including intensity difference, single-mode, and balanced homodyne detection. Furthermore, we analyze the QCRB associated with the QFI obtained for these states in all the aforementioned scenarios.\par

Our results demonstrate that the phase sensitivities for the Perelomov coherent input state in different detection schemes are better and would provide a more precise result than that of the Barut-Girardello coherent input state. The use of balanced homodyne detection techniques has been studied. The availability of an external phase reference can significantly enhance the performance of these input states, particularly in the unphysical limits where the transmission coefficients of the beam splitters approach $|\alpha|\rightarrow 1$ and $|\alpha'|\rightarrow 0$. This suggests that an external phase reference can improve the performance of these input states in quantum interferometry, potentially leading to improved measurement precision.\par 

Besides, entangled states offer a promising advantage in quantum metrology, particularly over Perelomov and Barut-Girardello coherent states. Using entanglement in interferometry, such as MZI, can improve phase sensitivity, a key factor in high-precision measurements. Coherent states are generally optimal for classical measurements, but in quantum metrology, entangled states can exceed these limits, enabling improvements in accuracy and sensitivity beyond the capabilities of coherent states. Our forthcoming work will probably explore specific mechanisms and test setups that exploit entanglement to improve performance in interferometric applications, highlighting the potential of quantum entanglement to achieve more accurate measurements.\\

{\bf Declaration of competing interest:}\par The authors declare that they have no known competing financial interests or personal relationships that could have appeared to influence the work reported in this paper.\\

{\bf Data availability:}\par No data was used for the research described in the article.

\end{document}